\documentclass[12pt]{article}
\usepackage{latexsym}
\usepackage{amsmath}
\usepackage{amssymb}
\usepackage{amsthm}
\usepackage{dsfont}
\usepackage{color}
\usepackage{verbatim}
\usepackage[sort&compress,numbers]{natbib}
\usepackage{graphicx}
\usepackage{epstopdf}
\usepackage{float}
\usepackage{ulem}

\newcommand{\Tr}{\mathop{\mathrm{Tr}}\nolimits}
\newcommand{\tg}{\mathop{\mathrm{tg}}\nolimits}
\newcommand{\ctg}{\mathop{\mathrm{ctg}}\nolimits}
\newcommand{\arctg}{\mathop{\mathrm{arctg}}\nolimits}
\newcommand{\diag}{\mathop{\mathrm{diag}}\nolimits}

\allowdisplaybreaks
\title{\bf\fontsize{19}{23}\selectfont Entanglement entropy of composite fermions realized by (deformed)
fermions vs. that of composite bosons}

\author{A. M. Gavrilik and Yu. A. Mishchenko}

\date{\small\it Bogolyubov Institute for Theoretical Physics of NAS of Ukraine,\\
14-b, Metrolohichna str., Kiev, 03680, Ukraine}

\begin{document}
\maketitle

\begin{abstract}
In our two preceding papers we studied bipartite composite boson (or
quasiboson) systems through their realization in terms of deformed oscillators.
Therein, the entanglement characteristics such as
the entanglement entropy and purity were found and expressed, for both
one-quasiboson and more complex states, through the parameter of deformation.
 In this work we initiate an analogous study of composite fermions
for two major cases: (i) ``boson + fermion'' composites;  (ii) ``deformed-boson + fermion''
composites. Both the entanglement entropy and purity of composite fermions are dealt with,
their dependence on the relevant parameters established, and for some particular
two- or three-mode cases depicted graphically.
In a few special cases the entanglement entropy turns out to be constant
$S_0=\ln 2$ (or $\ln 3$) or $S_0=0$, while in the rest of the cases
which we considered it varies between zero and $\ln 2$ (or $\ln 3$).
\end{abstract}



\section{Introduction}

Composite fermions (CFs) play significant role
in modern quantum physics. Suffice it to mention few distinct instances
of CFs: one taken from the domain of condensed matter physics, namely quasiparticles
involved in the theory of fractional quantum Hall effect~\cite{Jain2007Composite},
the other two -- baryons and pentaquarks -- belong to the realm of high energy
physics~\cite{Hadjimichef1998,Oh2004Penta,Browder2004Penta}.
  In this paper we focus on the composite fermions
with algebraic realization in two relatively simple cases: the first one
involves, as the constituents, pure fermion and pure boson,
while the second one concerns composites of pure fermion and
a deformed boson, the description of the latter being taken in rather general form.

Not less important are the composite bosons ({\it quasi-bosons}, cobosons) i.e.
non-elementary Bose-like systems or (quasi-)particles built from two or
more constituent particles. These are as well widely
encountered~\cite{Tichy_Rev,Avan,Hadjimichef1998,Perkins_2002,Moskalenko2000Bose,Bethe_Salpeter,Esquivel}
in modern quantum physics, both theoretical and experimental.
Among quasibosons there are excitons, cooperons, positronium, mesons,
diquarks or tetraquarks, odd-odd or even-even nuclei,  atoms, etc. In our preceding
works~\cite{GM_Entang,GM_Ent(En)} we focused on the case of bipartite
(two-component) composite bosons of two types: ``fermion + fermion'' and ``boson +
boson'' ones such that their creation and annihilation operators are given
through the typical ansatz,
\begin{equation}\label{ansatz}
A^{\dag}_{\alpha}= \sum_{\mu\nu}\Phi_{\alpha}^{\mu\nu} a^{\dag}_{\mu}b^{\dag}_{\nu},\
\quad A_{\alpha} = \sum_{\mu\nu} \overline{\Phi_{\alpha}^{\mu\nu}} b_{\nu} a_{\mu}  ,
\end{equation}
with $a^{\dag}_{\mu}$ and $b^{\dag}_{\nu}$ the creation
operators for the (distinguishable) constituents which can be taken
as either both fermionic or both bosonic. In~\cite{GKM2,GKM} it was
shown that the composite bosons of particular form (corresponding to
an appropriate matrices $\Phi_{\alpha}^{\mu\nu}$) can be realized,
in algebraic sense, by suitable deformed bosons (deformed oscillators).
Note that, with such realization in mind, one can then construct certain deformed
Bose gas model which serves for an effective description~\cite{GM2015muq-corr} of
the non-Bose like behavior of two-particle correlation
function intercepts of the pions $\pi^+$ and $\pi^-$ (also known as quark-antiquark
composites) produced in the experiments on Heavy ion collisions at RHIC.

An important concept used in quantum information theory, quantum communication and
teleportation~\cite{Horodecki,Tichy_Rev}, is the notion of entanglement or
quantum correlatedness between the constituents of composite particle or another
composite system. This concept was recently actively studied just in the context of
quasi-bosons~\cite{Law,Chudzicki,GM_Entang}. Among the measures or witnesses characterizing
the degree of entanglement, most widely used are the entanglement entropy and purity
(= inverse of the Schmidt number)~\cite{Horodecki,Tichy_Rev}.
The measures of entanglement between components of quasi-boson quantify to what extent
or accuracy the quasiboson approaches the properties of true boson~\cite{Law,Chudzicki,Ramanathan,Morimae}.

 For the composite bosons realizable by deformed oscillators it is
possible  to directly link~\cite{GM_Entang} the relevant parameter
of deformation with the entanglement characteristics of the composite boson.
 Then, the characteristics (or measures) of bipartite entanglement
with respect to $a$- and $b$-subsystems, see the ansatz (\ref{ansatz}), can be
found explicitly~\cite{GM_Entang}, and given through the deformation
parameter: for single composite boson, for multi-quasiboson states,
and for a coherent state corresponding to such quasi-bosons.

It is of importance to know what is the influence of system's energy
on the (variation of) quantum correlation and/or quantum statistics
properties of the system under study.
 The energy of a quasi-boson differs from the energy of the respective
ideal boson, and the difference (including quasiboson bound states energy)
essentially depends on the quasi-boson's entanglement,
and thus the latter clearly shows the deviation from bosonic behavior.
 Let us note in this context that the entanglement-energy relation is
relevant to quantum information research, quantum communication,
entanglement production~\cite{Weder}, quantum dissociation
processes~\cite{Esquivel}, particle addition or subtraction~\cite{Kurzynski,Bartley}.
In the case of
composite bosons (quasi-bosons) it was explored in~\cite{GM_Ent(En)},
and a number of interesting observations was obtained.

In this work we explore an alternative type of composites --
the composite fermions. Since the  entanglement entropy $S_{ent}$ is of primary
interest, we below, after appropriate analysis of the realization issue,
pay our main attention to finding the entanglement entropy $S_{ent}$
characterizing the composite fermion systems.
 Our treatment is performed for the one composite fermion states
(for comparison, the respective results for one  quasi-boson states
are also briefly sketched).
In some analogy with the case of quasi-bosons, we take the composite fermions as
bipartite systems realized in terms of mode-independent fermionic oscillators
(such independence is understood in fermionic sense).
Let us also note that the other entanglement measure -- purity --
will be considered, where appropriate, as well.

Let us emphasize that the investigation in this paper concern a single (or isolated)
 composite fermion states, not many-fermion system in some region of space.
  Accordingly, the considered entanglement and its entropy incorporate
 the two parties (two constituents) of the bipartite composite fermion.
  Just these features make our approach and analysis basically different
  from some recent works on the entanglement entropy of a system of
  free or composite fermions, see e.g.~\cite{Gioev2006Entanglement,Shao2014Entanglement},
  where the size of subsystem played basic role, and the very entanglement was viewed in
  a way fully different from ours.

The paper is organized as follows.
A sketch of main aspects on quasi-bosons is given in Sec.~\ref{sec:setup}.
  Major part contained in Sections \ref{sec:QFs}-\ref{sec:more_gen} deals with composite
fermions. First of all, we perform the analysis of algebraic realization
of composite fermions by means of (deformed) fermionic oscillators.
Then, the entanglement entropy of such (one-particle) CF states is explored in
Sec.~\ref{sec:2mod(nd)}-\ref{sec:more_gen}. Modified CFs --
those composed of fermion and deformed boson are analyzed in Sec.~\ref{sec:2-mod(def)}.
  The purity witness of bipartite entanglement of CF state
  is considered as well, see Section~\ref{sec:2mod(nd)}.
 The paper is concluded with short discussion of the essence of the obtained
results, of some implications and possible developments.


\section{Quasi-bosons formed as two-fermion (two-boson) composites}\label{sec:setup}

Let us recall main facts on the composite bosons realized by the
set of independent modes of deformed bosons (deformed oscillators),
given by the defining {\it deformation structure
function}~$\varphi(n)$. At the algebraic level the quasiboson
operators~$A_{\alpha}$, $A^\dag_{\alpha}$ and the number operator~$N_\alpha$
satisfy on the states the same relations as the corresponding deformed
oscillator creation/annihilation and occupation number operators:
\begin{align}
&A^\dag_\alpha A_\alpha = \varphi(N_\alpha),\\
&[A_\alpha,A^\dag_\beta] = \delta_{\alpha\beta} \bigl(\varphi(N_\alpha+1)-
\varphi(N_\alpha)\bigr),\\
&[N_\alpha,A^\dag_\beta] = \delta_{\alpha\beta} A^\dag_\beta,\quad [N_\alpha,A_\beta] =
 - \delta_{\alpha\beta} A_\beta .
\end{align}
Here Kronecker deltas reflect mode independence.
 Such realization implies~\cite{GKM,GKM2} that the structure
function~$\varphi(n)$ involves discrete deformation parameter $f$
and is quadratic in the occupation number~$n$ \ (set $\kappa=\pm 1$):
\begin{equation}\label{phi(n)}
\varphi(n)=\Bigl(1+\kappa \frac{f}{2}\Bigr)n - \kappa
\frac{f}{2}n^2,\quad f=\frac{2}{m},\ \
m=1,2,3,\ldots,
\end{equation}
while the matrices $\Phi_{\alpha}$ are of the form
\begin{equation}
\Phi_{\alpha} = U_1(d_a) \diag\Bigl\{0..0,\sqrt{f/2}\,
U_{\alpha}(m),0..0\Bigr\}U^{\dag}_2(d_b).\label{gen_solution}
\end{equation}
Note that the state of one composite boson,
\begin{equation}\label{1state}
|\Psi_{\alpha}\rangle \!=\! \sum_{\mu\nu}\Phi_{\alpha}^{\mu\nu}
|a_{\mu}\rangle\!\otimes\!|b_{\nu}\rangle,\quad |a_{\mu}\rangle\equiv
a^{\dag}_{\mu}|0\rangle,\ \ \ |b_{\nu}\rangle\equiv
b^{\dag}_{\nu}|0\rangle,
\end{equation}
is in general bipartite entangled relative to the states of two
constituent fermions (or two bosons).

  The extent of entanglement can be measured by the
well-known witnesses:  Schmidt rank, Schmidt number or its inverse
-- purity, entanglement entropy and concurrence~\cite{Tichy_Rev,Horodecki}.
As it was proven in~\cite{GM_Entang}, the entanglement entropy in
the case of one composite boson has the form
\begin{equation}
S_{\rm ent} = \ln(m) = \ln\frac2f .
\end{equation}
For the multi-quasibosonic states the respective extended results
were also obtained, see~\cite{GM_Entang,GM_Ent(En)}.

Purity $P$ is yet another popular witness of entanglement (see~\cite{Tichy_Rev,Horodecki}),
being the inverse $P=1/K$ of Schmidt number $K$.
Note that the {\it purity} is exploited in connection with the issue
of entanglement creation using scattering processes~\cite{Weder}
(for others contexts see~\cite{Kurzynski,McHugh}).
  For the {\it entangled system such as one quasiboson}, purity
is connected~\cite{GM_Entang} with the deformation parameter $m=\frac2f$
as follows:
\begin{equation}\label{pur_def}
P\!=\!\sum_k \lambda_k^4\!=\!\frac{1}{m},\ \ \ \text{or}\ \ \
P\!=\!\Tr (\rho_{\alpha}^{(a)})^2 \!=\! \Tr (\rho_{\alpha}^{(b)})^2
\!=\! \frac{1}{m}.
\end{equation}

\section{Composite fermions build as boson-fermion composites}\label{sec:QFs}

Now consider the composite fermions which are composed of pure boson
(or deformed boson) and pure fermion. The
CFs' creation, annihilation operators are given by the
same ``ansatz'' as in~(\ref{ansatz}),
where, this time, $a^{\dag}_{\mu}, a_{\mu}$ -- respectively creation and
annihilation operators for the constituent bosons (deformed or not) and
$b^{\dag}_{\nu}, b_{\nu}$ -- those for the constituent fermions, with
usual anticommutation relations for the latter.
 Suppose that different modes of deformed bosons are
independent.  Then, we obtain the following commutation and
defining relations for the operators of constituent bosons (deformed or not),
and fermions ($n^a_{\mu}$ denotes the particle number
operator for deformed bosons in $\mu$-mode):
  \begin{equation*} \left\{
\begin{aligned}
&a^{\dag}_{\mu}a_{\mu}=\chi(n^a_\mu);\\
&[a_{\mu},a^{\dag}_{\mu'}] = \delta_{\mu\mu'}\bigl(\chi(n^a_{\mu}+1) - \chi(n^a_{\mu})\bigr);
\quad [a^{\dag}_{\mu},a^{\dag}_{\mu'}]=0;\\
&[n^a_{\mu},a^{\dag}_{\mu}]=a^{\dag}_{\mu};
\end{aligned}
\right. \qquad\qquad \left\{
\begin{aligned}
&\{b_{\nu},b^{\dag}_{\nu'}\}=\delta_{\nu\nu'};\\
&\{b^{\dag}_{\nu},b^{\dag}_{\nu'}\}=0
\end{aligned}
\right.
 \end{equation*}
(here deformation structure function $\chi(n)$
corresponds to general case of deformed constituent boson; for non-deformed
i.e. usual boson $\chi(N)\equiv N$).

Remark that the normalization of the deformed boson
states, because of $a_{\mu} a^{\dag}_{\mu} = \chi(n^a_{\mu}+1)$,
implies $\chi(1)=1$. The CFs are supposed to be
independent (in the fermionic sense).
 We also suppose them to behave on the states as deformed particles with
structure function $\varphi(N)$. Having defined the particle number
operator for CFs as $N_{\alpha}=N_{\alpha}(A^\dag_\alpha
A_\alpha,A_\alpha A^\dag_\alpha,n^a_\mu,n^b_\nu)$ we infer the relations
\begin{align}
&\{A^\dag_\alpha,A^\dag_\beta\}\simeq0,\label{req1}\\
&A^\dag_\alpha A_\alpha \simeq \varphi(N_{\alpha}),\label{req2'}\\
&\{A_\alpha,A^\dag_\beta\} \simeq \delta_{\alpha\beta}[\varphi(N_{\alpha}+1)+\varphi(N_{\alpha})],\label{req2}\\
&[N_\alpha,A^\dag_\beta] \simeq \delta_{\alpha\beta}
A^\dag_\beta,\label{req3}
\end{align}
where the sign $\simeq$ (of weak equality) means equality on the
states, namely
 \begin{equation}
G\simeq G' \quad \Leftrightarrow \quad (G-G')
A^{\dag}_{\gamma_m}\ldots A^{\dag}_{\gamma_1}|0\rangle =0 \quad
\forall m\geq 0.\label{slabrav}
 \end{equation}
The first requirement (\ref{req1})
holds automatically and moreover in the strict sense:
\begin{equation}\label{indep(strict)}
\{A^{\dag}_{\alpha},A^{\dag}_{\beta}\} =
\Phi_{\alpha}^{\mu\nu}\Phi_{\beta}^{\mu'\nu'}
\{b^{\dag}_{\nu}a^{\dag}_{\mu},a^{\dag}_{\mu'}b^{\dag}_{\nu'}\} =
\Phi_{\alpha}^{\mu\nu}\Phi_{\beta}^{\mu'\nu'}
b^{\dag}_{\nu}b^{\dag}_{\nu'}[a^{\dag}_{\mu},a^{\dag}_{\mu'}] =0.
\end{equation}
As a consequence we come to the fermionic nilpotency property
 \begin{equation} \label{nilpot}
  (A^{\dag}_{\alpha})^2=0.
  \end{equation}
The next requirement eq.~(\ref{req2}) can be rewritten as a system of equations
 \begin{equation}\label{system1} \left\{
\begin{aligned}
&\{A_{\alpha},A^{\dag}_{\beta}\}A^{\dag}_{\gamma_m}\ldots A^{\dag}_{\gamma_1}|0\rangle = 0,\quad \alpha\neq \beta,\\
&\{A_{\alpha},A^{\dag}_{\alpha}\}(A^{\dag}_{\alpha})^m|0\rangle =
[\varphi(N_{\alpha}+1)+\varphi(N_{\alpha})](A^{\dag}_{\alpha})^m|0\rangle,\
\ \ m=\overline{0,1}.
\end{aligned}
\right.
 \end{equation}
Then, the anticommutator yields
\begin{multline}\label{expr1}
\{A_\alpha,A^\dag_\beta\} \!=\! \sum_\mu (\Phi_\beta
\Phi^\dag_\alpha)^{\mu\mu} [\chi(n^a_\mu\!+\!1)\!-\!\chi(n^a_\mu)] +
\sum_{\mu\mu'} (\Phi_\beta\Phi^\dag_\alpha)^{\mu'\mu} a^\dag_{\mu'}
a_\mu - \sum_{\mu\nu\nu'}
\overline{\Phi_\alpha^{\mu\nu}}\Phi^{\mu\nu'}_\beta
[\chi(n^a_\mu\!+\!1)\!-\!\chi(n^a_\mu)] b^\dag_{\nu'}b_\nu.
\end{multline}
Using the normalization of structural matrices
 \begin{equation}
  \label{orthonorm}
\Tr(\Phi_\beta \Phi^\dag_\alpha) = \delta_{\alpha\beta},
 \end{equation}
 we calculate (\ref{req2}) on the vacuum state:
 \begin{equation*}
[\chi(1)-\chi(0)]\delta_{\alpha\beta}|0\rangle =
\delta_{\alpha\beta}[\varphi(1)+\varphi(0)]|0\rangle \quad
\Rightarrow \quad \varphi(1)=\chi(1)=1 .
 \end{equation*}
For convenience, introduce the notation
  \begin{equation}\label{Delta_k}
    \Delta_k\chi(n^a_{\mu}) =
\sum_{l=0}^k (-1)^{k-l} C_k^l \chi(n^a_{\mu}+l),\ \ \ \ k=0,1,... ,
  \end{equation}
  where $C_k^l$ -- binomial coefficients.
The first few terms of the sequence $\{\Delta_k\chi\}$ are
 \begin{equation*}
 \Delta_0\chi(n^a_{\mu}) =
\chi(n^a_{\mu}),\ \ \ \Delta_1\chi(n^a_{\mu}) =
\chi(n^a_{\mu}\!+\!1)-\chi(n^a_{\mu}),\ \ \
\Delta_2\chi(n^a_{\mu})=\chi(n^a_{\mu}\!+\!2)-2\chi(n^a_{\mu}\!+\!1)+\chi(n^a_{\mu}).
  \end{equation*}
  Then, the following useful relations for $\Delta_k\chi$ do hold:
\begin{align*}
&[\Delta_k\chi(n^a_{\mu}),a^{\dag}_\mu] = a^{\dag}_\mu \Delta_{k+1}\chi(n^a_{\mu}),\\
&[\Delta_k\chi(n^a_{\mu}),A^{\dag}_\gamma] = \sum\nolimits_\nu
\Phi_\gamma^{\mu\nu} a^{\dag}_\mu b^{\dag}_\nu
\Delta_{k+1}\chi(n^a_\mu).
\end{align*}
Using (\ref{Delta_k}) the expression for the
anticommutator $\{A_{\alpha},A^{\dag}_{\beta}\}$ in (\ref{expr1})
 can be rewritten as
 \begin{equation*}
 \{A_{\alpha},A^{\dag}_{\beta}\} = \sum_\mu
(\Phi_{\beta}\Phi^{\dag}_{\alpha})^{\mu\mu}\Delta_1\chi(n^a_{\mu}) +
\sum_{\mu\mu'}
(\Phi_{\beta}\Phi^{\dag}_{\alpha})^{\mu'\mu}a^{\dag}_{\mu'}a_{\mu} -
\sum_{\mu\nu\nu'}
\overline{\Phi_{\alpha}^{\mu\nu}}\Phi^{\mu\nu'}_{\beta}b^{\dag}_{\nu'}b_{\nu}\Delta_1\chi(n^a_{\mu}).
   \end{equation*}
The latter for the case of {\it nondeformed constituent boson} ($\chi(n)\equiv n$)
  with the use of~(\ref{orthonorm}) reduces to
\begin{equation}
\{A_{\alpha},A^{\dag}_{\beta}\} = \delta_{\alpha\beta} +
\sum_{\mu\mu'}
(\Phi_{\beta}\Phi^{\dag}_{\alpha})^{\mu'\mu}a^{\dag}_{\mu'}a_{\mu} -
\sum_{\nu\nu'} (\Phi_{\alpha}^\dag\Phi_{\beta})^{\nu\nu'}
b^{\dag}_{\nu'}b_{\nu}.\label{expr1nondef}
\end{equation}
We also need the commutators
\begin{align*}
&[a_\mu,A^\dag_\gamma] = \sum\nolimits_\nu \Phi^{\mu\nu}_\gamma b^\dag_\nu \Delta_1\chi(n^a_\mu),\\
&[b^\dag_{\nu'} b_\nu \Delta_l\chi(n^a_\mu),A^\dag_\gamma] \!=\!
\sum_{\mu_1\nu_1} \Phi_\gamma^{\mu_1\nu_1}
a^\dag_{\mu_1}b^\dag_{\nu'}
\bigl[\delta_{\nu\nu_1}\bigl(\Delta_l\chi(n^a_\mu) \!+\!
\delta_{\mu\mu_1}\Delta_{l+1}\chi(n^a_\mu)\bigr) \!-\!
\delta_{\mu\mu_1}b^\dag_{\nu_1}b_\nu
\Delta_{l+1}\chi(n^a_\mu)\bigr].
\end{align*}
Besides, we calculate the commutator
\begin{multline}\label{commut1}
[\{A_\alpha,A^\dag_\beta\},A^\dag_\gamma] = \sum_{\mu\mu_1\nu_1}
\bigl[(\Phi_\beta \Phi^\dag_\alpha)^{\mu_1\mu}\Phi^{\mu\nu_1}_\gamma
- (\Phi_\gamma \Phi^\dag_\alpha)^{\mu_1\mu}
\Phi^{\mu\nu_1}_\beta\bigr] a^\dag_{\mu_1}b^{\dag}_{\nu_1}
\bigl(\Delta_1\chi(n^a_\mu) + \delta_{\mu\mu_1}\Delta_{2}\chi(n^a_\mu)\bigr) +\\
+ \sum_{\mu\nu\nu'\nu_1} \overline{\Phi_{\alpha}^{\mu\nu}}
\Phi^{\mu\nu'}_\beta \Phi_\gamma^{\mu\nu_1} a^\dag_\mu b^\dag_{\nu'}
b^\dag_{\nu_1}b_\nu \Delta_2\chi(n^a_\mu).
\end{multline}
Its nondeformed analog (when $\chi(n)\equiv n$) is
\begin{equation}\label{[[A,A]A]_nd}
[\{A_\alpha,A^\dag_\beta\},A^\dag_\gamma] = \sum_{\mu\nu}
\bigl(\Phi_\beta \Phi^\dag_\alpha\Phi_\gamma - \Phi_\gamma
\Phi^\dag_\alpha \Phi_\beta\bigr)^{\mu\nu} a^\dag_\mu b^\dag_\nu.
\end{equation}
Setting in~(\ref{commut1}) $\alpha=\beta=\gamma$ we get
 \begin{equation*}
[\{A_\alpha,A^\dag_\alpha\},A^{\dag}_{\alpha}] =
0\quad\Rightarrow\quad
\{A_\alpha,A^\dag_\alpha\}(A^{\dag}_{\alpha})^m|0\rangle =
(A^{\dag}_{\alpha})^m|0\rangle,\quad m=\overline{0,1}.
  \end{equation*}
 On the other hand, using the relation (\ref{req3}) (which needs
a verification afterwards) we calculate the
corresponding r.h.s. according to~(\ref{req2}):
 \begin{equation*}
[\varphi(N_\alpha+1)+\varphi(N_\alpha)](A^\dag_\alpha)^m|0\rangle =
[\varphi(m+1)+\varphi(m)](A^{\dag}_{\alpha})^m|0\rangle,\quad
m=\overline{0,1}.
  \end{equation*}
The second requirement in system (\ref{system1}) now takes the form
(note that there should be $\varphi(0)=0$)
\begin{equation*}
\varphi(m+1)+\varphi(m)=\chi(1)=1,\ \ \
m=\overline{0,1}\quad\Rightarrow\quad \varphi(2)=0.
\end{equation*}
  This is similar to fermionic structure function.
  Then, the r.h.s. of~(\ref{req2}) on the states commutes with $A^\dag_\gamma$
and therefore
\begin{equation*}
[\varphi(N_{\alpha}+1)+\varphi(N_{\alpha}),A^\dag_\gamma] \simeq
\delta_{\alpha\gamma} A^\dag_\gamma
\bigl(\varphi(N_\alpha+2)-\varphi(N_\alpha)\bigr) =0.
\end{equation*}
Thus, considering~(\ref{req2}) on the one-CF states we
obtain the equation
 \begin{equation}
 \label{req2_chi(2)} (\Phi_\beta\Phi^\dag_\alpha\Phi_\gamma)^{\mu\nu}
- (\Phi_\gamma\Phi^\dag_\alpha\Phi_\beta)^{\mu\nu} +
\bigl[(\Phi_\beta\Phi^\dag_\alpha)^{\mu\mu} \Phi_\gamma^{\mu\nu} -
(\Phi_\gamma\Phi^\dag_\alpha)^{\mu\mu}\Phi^{\mu\nu}_\beta\bigr]
\bigl(\chi(2)-2\bigr)=0.
  \end{equation}
  In the case of non-deformed constituent boson this relation
  due to~(\ref{[[A,A]A]_nd}) yields
$[\{A_\alpha,A^\dag_\beta\},A^\dag_\gamma] = 0$, and thus the
realization conditions on matrices $\Phi_\alpha$, see~(\ref{req2}),
(\ref{orthonorm}), take the form
\begin{equation}\label{nondef}
\left\{
\begin{aligned}
&\Tr(\Phi_\beta\Phi^\dag_\alpha) = \delta_{\alpha\beta},\\
&\Phi_\beta\Phi^\dag_\alpha\Phi_\gamma -
\Phi_\gamma\Phi^\dag_\alpha\Phi_\beta=0.
\end{aligned}
\right.
\end{equation}

 It is not difficult to calculate the following double (anti)commutator
\begin{multline*}
\{[\{A_\alpha,A^\dag_\beta\},A^\dag_{\gamma_1}],A^\dag_{\gamma_2}\}
= \sum_{\mu\mu_1\nu_1\nu_2}
\bigl[(\Phi_\beta\Phi^\dag_\alpha)^{\mu_1\mu}
\Phi^{\mu\nu_1}_{\gamma_1} \Phi_{\gamma_2}^{\mu\nu_2} -
(\Phi_{\gamma_1}
\Phi^\dag_\alpha)^{\mu_1\mu}\Phi^{\mu\nu_1}_\beta\Phi_{\gamma_2}^{\mu\nu_2}
+  (\Phi_{\gamma_2}\Phi_\alpha^\dag)^{\mu_1\mu}\Phi^{\mu\nu_1}_\beta \Phi_{\gamma_1}^{\mu\nu_2}\bigr] \cdot\\
\cdot a^\dag_\mu a^\dag_{\mu_1} b^{\dag}_{\nu_1}b^{\dag}_{\nu_2}
\bigl(\Delta_2\chi(n^a_\mu) +  \delta_{\mu\mu_1}
\Delta_3\chi(n^a_\mu)\bigr) - \sum_{\mu\nu\nu'\nu_1\nu_2}
\overline{\Phi_{\alpha}^{\mu\nu}}\Phi^{\mu\nu'}_{\beta}\Phi_{\gamma_1}^{\mu\nu_1}\Phi_{\gamma_2}^{\mu\nu_2}
(a^{\dag}_\mu)^2 b^\dag_{\nu'}b^\dag_{\nu_1} b^\dag_{\nu_2}b_\nu
\Delta_3\chi(n^a_\mu).
\end{multline*}
and likewise the higher (anti)commutator
$[\{[\{A_\alpha,A^\dag_\beta\},A^\dag_{\gamma_1}],A^\dag_{\gamma_2}\},A^\dag_{\gamma_3}]$
 (we omit the latter).

\section{The cases of one and two composite fermion modes}\label{sec:2mod(nd)}

In the case of single CF mode $\alpha$, it is enough to
consider the realization conditions (\ref{req1})-(\ref{req3}) on the
vacuum and on the one-CF state. This yields $\Tr(\Phi_\alpha
\Phi^\dag_\alpha) = \varphi(1)+\varphi(2) = 1$. Its general solution
can be written in the form of singular value decomposition (linked with
Schmidt decomposition)
\begin{equation*}
\Phi_\alpha = U_\alpha
\diag\{\lambda^{(\alpha)}_1,\lambda^{(\alpha)}_2,...\} V^\dag_\alpha
\end{equation*}
with real non-negative $\lambda^{(\alpha)}_i$ written in the descending order
such that $\sum_i (\lambda^{(\alpha)}_i)^2 = 1$ (no summation over $\alpha$), and an
arbitrary unitary matrices $U_\alpha$, $V_\alpha$. Entanglement entropy within a
composite fermion (i.e. between its constituents) viewed
as bipartite system equals~\cite{Horodecki,Tichy_Rev}
\begin{equation}\label{char3}
S_{\rm entang}= -\sum_i (\lambda^{(\alpha)}_i)^2 \ln
(\lambda^{(\alpha)}_i)^2.
\end{equation}

When just two CF modes $\alpha=1,\,2$ are dealt with, in
the case of a non-deformed constituent boson system~(\ref{nondef})
reduces to the set of independent equations
\begin{align}
&\Tr(\Phi_i\Phi^\dag_i) = 1,\ \ i=\overline{1,2}, \quad
   \Tr(\Phi_1\Phi^\dag_2) =0;\label{eq1nd}\\
&\Phi_1\Phi^\dag_1\Phi_2 - \Phi_2\Phi^\dag_1\Phi_1=0;\label{eq2nd}\\
&\Phi_1\Phi^\dag_2\Phi_2 - \Phi_2\Phi^\dag_2\Phi_1=0.\label{eq3nd}
\end{align}
To solve these we use the singular value decomposition for
$\Phi_1$ and make the replacement $\Phi_2 \to \tilde{\Phi}_2$:
\begin{equation}\label{repl}
\Phi_1 = U_1 D_1 V^\dag_1,\quad \Phi_2 = U_1 \tilde{\Phi}_2 V^\dag_1,
\end{equation}
where $D_1=\diag\{\lambda^{(1)}_1,\lambda^{(1)}_2,...\}$ is some diagonal matrix,
$\lambda^{(1)}_i\ge \lambda^{(1)}_j$ for $i<j$, and $U_1$, $V_1$ are unitary matrices.
 Then the system~(\ref{eq1nd})-(\ref{eq3nd}) is presented as
\begin{align}
&\Tr(D_1^2) \equiv \sum\nolimits_i (\lambda^{(1)}_i)^2 =
\Tr(\tilde{\Phi}_2\tilde{\Phi}^\dag_2) = 1,
\quad \Tr(D_1\tilde{\Phi}^\dag_2) =0;\label{eq1'nd}\\
&D_1^2\tilde{\Phi}_2 - \tilde{\Phi}_2 D_1^2 = 0;\label{eq2'nd}\\
&D_1\tilde{\Phi}^\dag_2 \tilde{\Phi}_2 - \tilde{\Phi}_2
\tilde{\Phi}^\dag_2 D_1=0.\label{eq3'nd}
\end{align}
Let us point out one particular solution
of~(\ref{eq1'nd})-(\ref{eq3'nd}).
 For this, we put $D_1\sim E$ that yields
\begin{equation}\label{norm_sol}
\tilde{\Phi}^\dag_2 \tilde{\Phi}_2 = \tilde{\Phi}_2
\tilde{\Phi}^\dag_2,\quad \Tr \tilde{\Phi}_2 = 0,\quad
\Tr(\tilde{\Phi}_2\tilde{\Phi}^\dag_2)=1.
\end{equation}
So, $\tilde{\Phi}_2$ is proportional to normal (i.e. commuting with
its own conjugate) traceless matrix.

Next we restrict ourselves to the case when the constituent boson
and constituent fermion can be in two modes, that is $\mu=1,\,2$ and $\nu=1,\,2$.
Then matrices $D_1$ and $\tilde{\Phi}_2$ are presented as
\begin{equation}
\tilde{\Phi}_2= \left(
\begin{array}{cc}
    \phi^{(2)}_{11} & \phi^{(2)}_{12}\\
    \phi^{(2)}_{21} & \phi^{(2)}_{22}
\end{array}
\right) = e^{i\eta} \tilde{U} D_2 \tilde{V}^\dag,\quad
D_\alpha = \left(
\begin{array}{cc}
    \lambda^{(\alpha)}_1 & 0 \\
    0 & \lambda^{(\alpha)}_2
\end{array}
\right),
\ \ \alpha=\overline{1,2},\ \ \ \tilde{U},\tilde{V}\in SU(2),\label{Phi2tilde-def}
\end{equation}
with $\tilde{U} = \Bigl(
\begin{array}{cc}
    \tilde{u}_1 & \tilde{u}_2 \\
    -\overline{\tilde{u}_2} & \overline{\tilde{u}_1}
\end{array}
\Bigr)$,
$\tilde{V} = \Bigl(
\begin{array}{cc}
    \tilde{v}_1 & \tilde{v}_2 \\
    -\overline{\tilde{v}_2} & \overline{\tilde{v}_1}
\end{array}
\Bigr)$, $|\tilde{u}_1|^2+|\tilde{u}_2|^2 = |\tilde{v}_1|^2+|\tilde{v}_2|^2 = 1$,
$\lambda^{(\alpha)}_i\ge 0$.
Eqs.~(\ref{eq1'nd}) for the traces are rewritten in the form
\begin{align}
&\Tr(D_\alpha^2) \equiv (\lambda^{(\alpha)}_1)^2 + (\lambda^{(\alpha)}_2)^2 = 1,\ \ \alpha=\overline{1,2},\nonumber\\
&\Tr(D_1\tilde{\Phi}^\dag_2) = e^{-i\eta} \bigl(\lambda^{(1)}_1 (\lambda^{(2)}_1 \overline{\tilde{u}_1}\tilde{v}_1
+ \lambda^{(2)}_2 \overline{\tilde{u}_2}\tilde{v}_2)
+ \lambda^{(1)}_2 (\lambda^{(2)}_1 \tilde{u}_2\overline{\tilde{v}_2}
+ \lambda^{(2)}_2 \tilde{u}_1\overline{\tilde{v}_1})\bigr)  =0.\label{eq1''nd}
\end{align}
Equation~(\ref{eq2'nd}) yields the system
\begin{equation}\label{eq2''nd}
((\lambda^{(1)}_1)^2-(\lambda^{(1)}_2)^2) \phi^{(2)}_{ij} = 0,\quad i\ne j, \ \ i,j=\overline{1,2}.
\end{equation}
Analogously, from~(\ref{eq3'nd}) we obtain:
\begin{align}
&D_1\tilde{\Phi}^\dag_2 \tilde{\Phi}_2 - \tilde{\Phi}_2\tilde{\Phi}^\dag_2 D_1
= ((\lambda^{(2)}_1)^2-(\lambda^{(2)}_2)^2)  \left(
\begin{array}{cc}
    \lambda^{(1)}_1 (|\tilde{v}_1|^2-|\tilde{u}_1|^2) & \lambda^{(1)}_2 \tilde{u}_1\tilde{u}_2
    - \lambda^{(1)}_1 \tilde{v}_1\tilde{v}_2\\
    \lambda^{(1)}_1 \overline{\tilde{u}_1}\,\overline{\tilde{u}_2} - \lambda^{(1)}_2
    \overline{\tilde{v}_1}\,\overline{\tilde{v}_2} & \lambda^{(1)}_2 (|\tilde{u}_1|^2-|\tilde{v}_1|^2)
\end{array}\right) = 0\nonumber\\
&\Longrightarrow\qquad \lambda^{(2)}_1 = \lambda^{(2)}_2 \quad \text{or} \quad
|\tilde{u}_1|=|\tilde{v}_1|,\ \ \ \lambda^{(1)}_2 \tilde{u}_1\tilde{u}_2
    = \lambda^{(1)}_1 \tilde{v}_1\tilde{v}_2. \label{eq3''nd}
\end{align}

\begin{figure}[h]
\centering
\includegraphics[width=0.49\columnwidth]{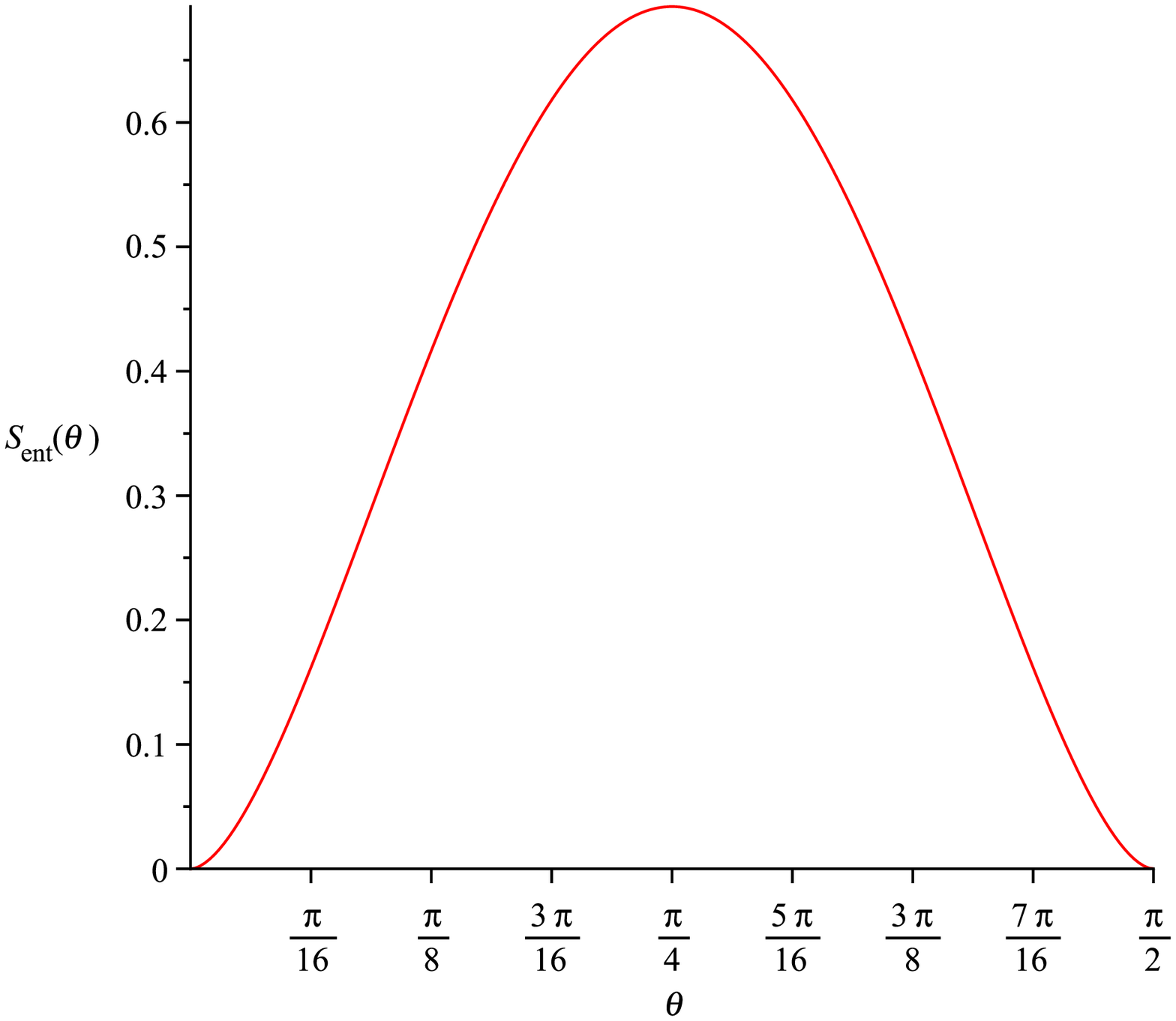}
\includegraphics[width=0.45\columnwidth]{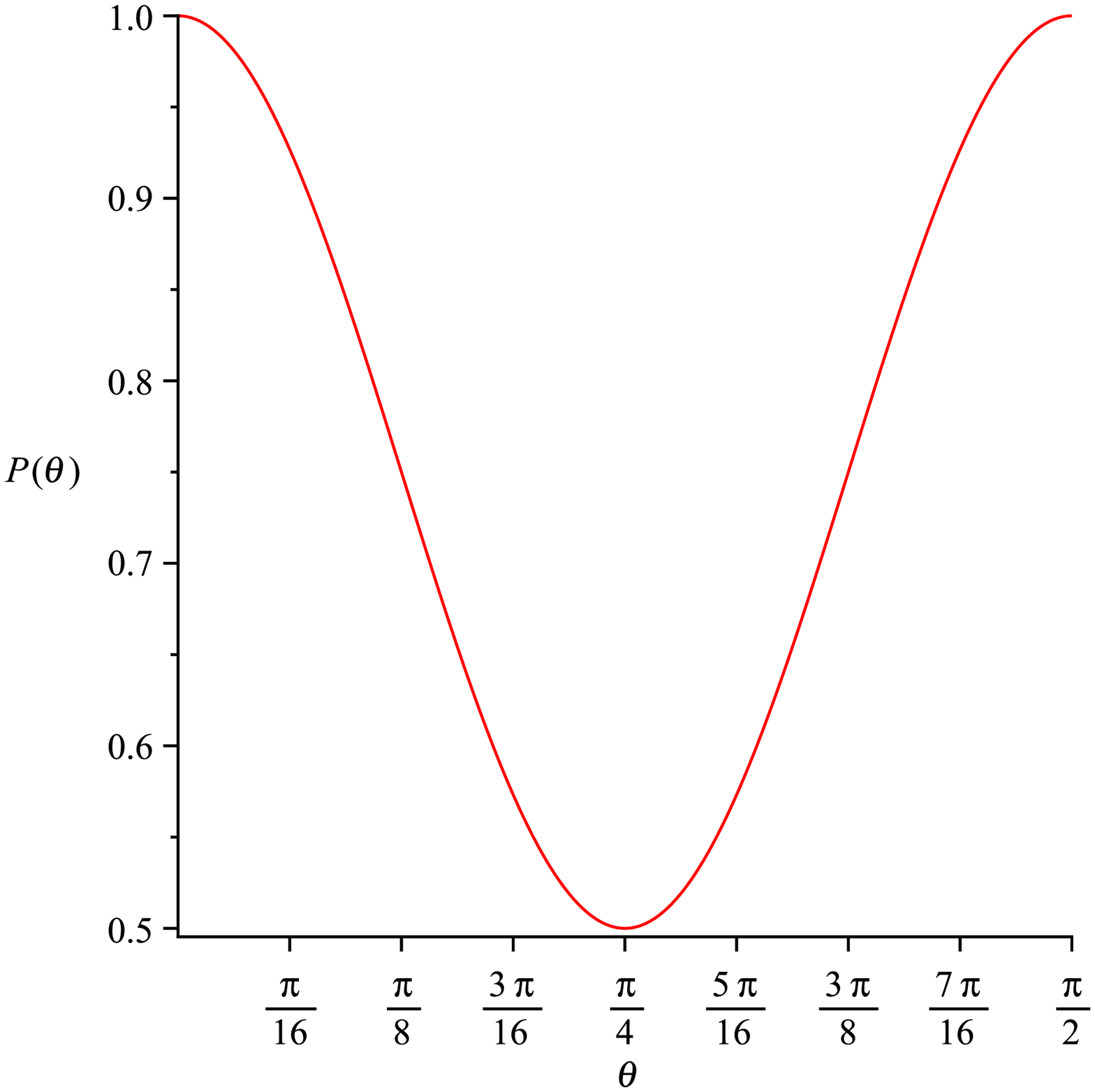}
\caption{Entanglement entropy $S_{\rm ent}$ and purity $P$ versus parameter
$\theta$ where $\cos\theta = \lambda^{(1)}_1$.
The conventional ordering $\lambda^{(1)}_1\ge \lambda^{(1)}_2$ corresponds
to $0\le\theta\le \frac{\pi}{4}$.}
\label{fig2}
\end{figure}

\noindent$\bullet$ \ If $\lambda^{(1)}_1 \ne \lambda^{(1)}_2$ eq.~(\ref{eq2''nd}) yields
$\phi^{(2)}_{12} = \phi^{(2)}_{21} = 0$, so that
using~(\ref{eq1''nd}) we obtain $\tilde{\Phi}_2 = \diag\{\phi^{(2)}_{11},\phi^{(2)}_{22}\}
= e^{i\eta'} \diag\{\lambda^{(1)}_2,-\lambda^{(1)}_1\}$.
 As result, the {\it entanglement entropy within the composite fermion},
realized by fermion, in each of the two modes equals
\begin{align}
&S_{\rm ent}|_{\alpha=1,2} = -(\lambda^{(1)}_1)^2 \ln (\lambda^{(1)}_1)^2 - (1-(\lambda^{(1)}_1)^2)
\ln (1-(\lambda^{(1)}_1)^2) = S_2(\theta),\label{S_2(a)}\\
&S_2(\theta)\equiv -\sin^2\theta \ln\sin^2\theta
- \cos^2\theta\ln \cos^2\theta,\quad \lambda^{(1)}_1=\cos\theta,\ \
0<\lambda^{(1)}_1<1, \ \ 0<\theta<\frac{\pi}{4}.
\label{S_2-def}
\end{align}
For illustration, this result is pictured in Fig.~\ref{fig2}
(left). It shows that the entanglement entropy
ranges from the value $S_{\rm ent}=0$ (at $\lambda^{(1)}_1=1$) to the value $S_{\rm ent}= \ln 2$
(at $\lambda^{(1)}_1=1/\sqrt{2}$) with $\ln 2$ being the maximum.
For comparison, let us also give the expression for the other entanglement measure
-- {\it purity} $P$ of the CF state,
\begin{equation}
P|_{\alpha=1,2}\equiv \sum\nolimits_i |\lambda^\alpha_i|^4 = (\lambda^{(1)}_1)^4
+ (1-(\lambda^{(1)}_1)^2)^2 = \frac14 (3+\cos 4\theta),\quad
0<\lambda^{(1)}_1<1,\ \ \ 0<\theta<\frac{\pi}{2}.
\end{equation}
The purity ranges from $P=1/2$ (at $\lambda^{(1)}_1=1/\sqrt{2}$) to $P=1$
(at $\lambda^{(1)}_1=1$), see~Fig.~\ref{fig2} (right).

\noindent$\bullet$ \ In the case of $\lambda^{(1)}_1 = \lambda^{(1)}_2$, from~(\ref{eq1''nd})
and (\ref{eq3''nd}) we have $\lambda^{(2)}_1 = \lambda^{(2)}_2$
since otherwise, i.e. for $\lambda^{(2)}_1 \ne \lambda^{(2)}_2$, in view of (\ref{eq3''nd})
we have $\Tr(D_1\tilde{\Phi}^\dag_2) = e^{-i\eta} \lambda^{(1)}_1 \bigl(\lambda^{(2)}_1
\frac{\overline{\tilde{u}_1}}{\overline{\tilde{v}_1}} + \lambda^{(2)}_2
\frac{\tilde{u}_1}{\tilde{v}_1}\bigr) \ne0$ that contradicts~(\ref{eq1''nd}).
In this case according to~(\ref{Phi2tilde-def}) we have
\begin{equation*}
\tilde{\Phi}_2 = e^{i\eta} \lambda^{(2)}_1 \tilde{U}, \quad \Tr \tilde{U} =0,
\end{equation*}
and the respective Schmidt coefficient squared $(\lambda^{(\alpha)}_i)^2$, $\alpha,i=\overline{1,2}$,
is equal to $1/2$. So, the entanglement entropy within the composite fermion in each
of the two modes $\alpha=1$ or $2$ is $S_{\rm ent}=\ln 2$, that is
the constant which coincides with maximal value for the case of (\ref{S_2(a)}).

\section{Composite fermions as composites of fermion and deformed boson: two-mode case}
\label{sec:2-mod(def)}

Let us go over to the two-mode case ($\alpha=1,\,2$) of
CF when it is composed of usual fermion and, say,
$\chi$-{deformed} boson. In this case the specifics of two modes for the
CFs implies that it is again sufficient to consider realization
conditions~(\ref{req2'})-(\ref{req3}) on the vacuum and
one-CF states. Indeed, for single non-zero
two-CF state $A_1^\dag A_2^\dag|0\rangle$, implying that
realization conditions ~(\ref{req2'})-(\ref{req3}) hold on the
vacuum and one-CF states, we have
\begin{align*}
&(A_1^\dag A_1-\varphi(N_1)) A_1^\dag A_2^\dag|0\rangle = -A_2^\dag
(A_1^\dag A_1-\varphi(N_1))
A_1^\dag |0\rangle = 0;\\
&(A_1 A_1^\dag-\varphi(N_1+1)) A_1^\dag A_2^\dag|0\rangle = A_2^\dag
\varphi(N_1+1)
A_1^\dag|0\rangle = 0;\\
&\{A_1,A_2^\dag\} A_1^\dag A_2^\dag|0\rangle = A_2^\dag A_1 A_1^\dag
A_2^\dag|0\rangle = (A_2^\dag)^2 A_1 A_1^\dag|0\rangle =0.
\end{align*}
The corresponding (to one-CF states) realization
condition~(\ref{req2_chi(2)}) then reduces to the following two
independent equations (denote $\delta\chi_2 \equiv \chi(2)\!-\!2$):
\begin{align}
&\Phi_1 \Phi_1^\dag \Phi_2 - \Phi_2 \Phi_1^\dag \Phi_1 + \delta\chi_2
\bigl[\diag\{(\Phi_1\Phi_1^\dag)^{\mu\mu}\} \Phi_2
- \diag\{(\Phi_2\Phi_1^\dag)^{\mu\mu}\} \Phi_1\bigr] =0,\label{eq1def}\\
&\Phi_1 \Phi_2^\dag \Phi_2 - \Phi_2 \Phi_2^\dag \Phi_1 +
\delta\chi_2 \bigl[\diag\{(\Phi_1\Phi_2^\dag)^{\mu\mu}\} \Phi_2 -
\diag\{(\Phi_2\Phi_2^\dag)^{\mu\mu}\} \Phi_1\bigr] =0.\label{eq2def}
\end{align}
Performing the replacement~(\ref{repl}) as in the case of
non-deformed constituent boson we arrive at the following system of
equations equivalent to~(\ref{eq1def}), (\ref{eq2def}), but now given in terms
of $D_1$ and $\tilde{\Phi}_2$:
\begin{align}
&D_1^2 \tilde{\Phi}_2 - \tilde{\Phi}_2
D_1^2 + \delta\chi_2 \bigl[U_1^\dag \diag\{(U_1
D_1^2 U_1^\dag)^{\mu\mu}\}U_1 \tilde{\Phi}_2
- U_1^\dag \diag\{(U_1 \tilde{\Phi}_2 D_1U_1^\dag)^{\mu\mu}\} U_1 D_1\bigr] =0,\label{eq1'def}\\
&D_1 \tilde{\Phi}_2^\dag \tilde{\Phi}_2 - \tilde{\Phi}_2
\tilde{\Phi}_2^\dag D_1 + \delta\chi_2 \bigl[U_1^\dag\diag\{(U_1 D_1
\tilde{\Phi}_2^\dag U_1^\dag)^{\mu\mu}\} U_1 \tilde{\Phi}_2 -
U_1^\dag \diag\{(U_1 \tilde{\Phi}_2\tilde{\Phi}_2^\dag
U_1^\dag)^{\mu\mu}\} U_1 D_1\bigr] =0 .   \label{eq2'def}
\end{align}
 To find the Schmidt coefficients $\lambda^\alpha_i$, $\alpha=1,\,2$,
contained in the definition of entanglement entropy it may be convenient
to deal with the variables $X_\alpha=\Phi_\alpha \Phi_\alpha^\dag$, $Y=\Phi_2\Phi_1^\dag$,
since $|\lambda^\alpha_i|^2$ are the eigenvalues of $X_\alpha$.
Multiplying (\ref{eq1def}) and (\ref{eq2def}) y $\Phi_1$ from the
right we obtain the equations
\begin{align*}
&X_1 Y-Y X_1 + \delta\chi_2 [\diag\{X_1^{\mu\mu}\} Y - \diag\{Y^{\mu\mu}\} X_1] = 0,\\
&Y^\dag Y - X_2 X_1 +\delta\chi_2 [\diag\{\overline{Y^{\mu\mu}}\} Y
- \diag\{X_2^{\mu\mu}\} X_1] = 0
\end{align*}
which for nondegenerate $\Phi_1$ are equivalent to (\ref{eq1def}), (\ref{eq2def}).

Restricting ourselves to the case of two modes $\mu,\nu=1,\,2$
of the constituents, without loss of generality we take $U_1\in SU(2)$.
 Using the parametrization: $U_1 = \Bigl(
\begin{array}{cc}
    u & v \\
    -\overline{v} & \overline{u}
\end{array}
\Bigr)$, $|u|^2+|v|^2=1$, and the identity
\begin{equation}
U_1^\dag \diag\{(U_1 X U_1^\dag)^{\mu\mu}\} U_1 = \frac12 X + \frac12 RXR, \quad
R = \Bigl(
\begin{array}{cc}
    |u|^2\!-\!|v|^2 & 2\overline{u}v \\
    2u\overline{v} & |v|^2\!-\!|u|^2
\end{array}\Bigr),
\end{equation}
we rewrite equations~(\ref{eq1'def}) and (\ref{eq2'def}) respectively as
\begin{align}
&\frac{\chi(2)}{2} (D_1^2 \tilde{\Phi}_2 - \tilde{\Phi}_2 D_1^2) + \frac{\delta\chi_2}{2}
\bigl(R D_1^2 R \tilde{\Phi}_2 - R \tilde{\Phi}_2 D_1 R D_1\bigr) =0,\label{eq1''def}\\
&\frac{\chi(2)}{2} (D_1 \tilde{\Phi}_2^\dag \tilde{\Phi}_2 - \tilde{\Phi}_2
\tilde{\Phi}_2^\dag D_1) + \frac{\delta\chi_2}{2} \bigl(R D_1 \tilde{\Phi}_2^\dag R \tilde{\Phi}_2
- R \tilde{\Phi}_2\tilde{\Phi}_2^\dag R D_1\bigr) =0 .   \label{eq2''def}
\end{align}
Taking into account three-dimensionality of the subspace of matrices
$\tilde{\Phi}^\dag_2$ satisfying the orthogonality condition $\Tr(D_1\tilde{\Phi}^\dag_2) =0$
we look for the solution of (\ref{eq1''def})-(\ref{eq2''def}) as the linear combination
of the following basis elements:
\begin{equation}
\tilde{\Phi}_2 = x_1 \Bigl(
\begin{array}{cc}
    \lambda^{(1)}_2 & 0 \\
    0 & -\lambda^{(1)}_1
\end{array}\Bigr) + x_2 \Bigl(
\begin{array}{cc}
    0 & \varkappa \lambda^{(1)}_1\\
    \overline{\varkappa}\lambda^{(1)}_2 & 0
\end{array}\Bigr) + x_3 \Bigl(
\begin{array}{cc}
    0 & -\varkappa \lambda^{(1)}_2 \\
    \overline{\varkappa} \lambda^{(1)}_1 & 0
\end{array}\Bigr),\ \ \ \varkappa = e^{i(\arg v -\arg u)}.
\end{equation}
 Then, after some calculation equation (\ref{eq1''def})
reduces to the system of linear (in $x_1,x_2,x_3$) equations
\begin{align}
&2\delta\chi_2 |u|^2|v|^2 x_1 - \delta\chi_2 |u||v| (|u|^2-|v|^2) x_2 = 0,\label{eq1(1)}\\
&- \delta\chi_2 |u||v| (|u|^2\!-\!|v|^2) x_1 \!+\! \frac12 \bigl(\chi(2) ((\lambda^{(1)}_1)^2\!-\!(\lambda^{(1)}_2)^2)^2
\!+\! \delta\chi_2 (|u|^2\!-\!|v|^2)^2\bigr)x_2 +\nonumber\\
&\hspace{2cm}+\chi(2)\lambda^{(1)}_1 \lambda^{(1)}_2 ((\lambda^{(1)}_2)^2\!-\!(\lambda^{(1)}_1)^2)x_3 \!=\! 0,\label{eq1(2)}\\
&\chi(2)\lambda^{(1)}_1 \lambda^{(1)}_2 ((\lambda^{(1)}_2)^2-(\lambda^{(1)}_1)^2) x_2 - \frac12 \bigl(\chi(2)
((\lambda^{(1)}_1)^2-(\lambda^{(1)}_2)^2)^2 - \delta\chi_2\bigr) x_3 =0.\label{eq1(3)}
\end{align}
For the existence of a nontrivial solution, the determinant of this system should be
zero, i.e.
\begin{equation*}
\det\bigl(...\bigr) = -\chi(2)(\chi(2)-2) |u|^2|v|^2
((\lambda^{(1)}_1)^2-(\lambda^{(1)}_2)^2)^2 = 0.
\end{equation*}
That is possible in the following cases:
\par\noindent{\bf a)} $\chi(2)=2$. Then eqs.~(\ref{eq1'def}), (\ref{eq2'def})
reduce to non-deformed eqs.~(\ref{eq2'nd}), (\ref{eq3'nd}) which
were already considered.
\par\noindent{\bf b)} $\chi(2)=0$ or $\lambda^{(1)}_1=\lambda^{(1)}_2$ at $\chi(2)\ne 2$.
Though the situation $\lambda^{(1)}_1=\lambda^{(1)}_2$ is qualitatively different,
in this case the solution of (\ref{eq1(1)})-(\ref{eq1(3)}) is given uniformly, namely
\begin{equation}\label{Phi2(b)}
\tilde{\Phi}_2 = \kappa R \left(
\begin{array}{cc}
    \lambda^{(1)}_2 & 0 \\
    0 & \lambda^{(1)}_1
\end{array}
\right), \quad |\kappa| = 1, \ \ (\lambda^{(1)}_1)^2+(\lambda^{(1)}_2)^2=1,
\end{equation}
{\it yielding the entanglement entropy}
\begin{equation}\label{S_ent0}
S_{\rm ent}|_{\alpha=1,2} = -(\lambda^{(1)}_1)^2 \ln (\lambda^{(1)}_1)^2
- (1-(\lambda^{(1)}_1)^2) \ln (1-(\lambda^{(1)}_1)^2)\
 = S_2(\theta),\quad 0\le\theta\le\frac{\pi}{4}
\end{equation}
(to be compared with (\ref{S_2(a)})-(\ref{S_2-def})) which for $\lambda^{(1)}_1=\lambda^{(1)}_2=\frac{1}{\sqrt{2}}$
yields $S_{\rm ent}|_{\alpha=1,2}=\ln 2$.
\par\noindent{\bf c)} $uv=0$ while $\chi(2)$ is unrestricted, and
$\lambda^{(1)}_1\ne\lambda^{(1)}_2$.
 Equation~(\ref{eq1''def}) takes the form
\begin{equation*}
\bigl(\chi(2) (\lambda^{(1)}_1)^2-1\bigr) \phi^{(2)}_{12} = 0,
\quad \bigl(\chi(2) (\lambda^{(1)}_2)^2-1\bigr) \phi^{(2)}_{21} = 0.
\end{equation*}
Eq.~(\ref{eq2''def}) e.g. with $\phi^{(2)}_{21}=0$ reduces to
\begin{equation*}
\left(
\begin{array}{cc}
    -(\chi(2)-1) \lambda^{(1)}_1 |\phi^{(2)}_{12}|^2 & \frac12\chi(2)
    (\lambda^{(1)}_1 \overline{\phi^{(2)}_{11}} - \lambda^{(1)}_2 \overline{\phi^{(2)}_{22}}) \phi^{(2)}_{12} \\
    (\lambda^{(1)}_2 \phi^{(2)}_{11} - \lambda^{(1)}_1 \phi^{(2)}_{22}) \overline{\phi^{(2)}_{12}}
    &\lambda^{(1)}_2 |\phi^{(2)}_{12}|^2
\end{array}\right) = 0.
\end{equation*}
So, there are two solutions:
\begin{itemize}
\item $\tilde{\Phi}_2 = \kappa \diag\bigl\{\lambda^{(1)}_2,
-\lambda^{(1)}_1\bigr\}$, $(\lambda^{(1)}_1)^2
+ (\lambda^{(1)}_2)^2=1$, $|\kappa|=1$, so that \ \
$S_{\rm ent}|_{\alpha=1,2} = S_2(\theta)$, $\lambda^{(1)}_1=\cos\theta$,
$0\le\theta\le\frac{\pi}{4}$.
Note, this result coincides with (\ref{S_ent0}), see also eq.~(\ref{S_2(a)}) and Fig.~\ref{fig2}.
\item If $\chi(2)=1$, there appears the additional solution $\tilde{\Phi}_2 = \left(
\begin{array}{cc}
    0 & \phi^{(2)}_{12} \\
    0 & 0
\end{array}\right)$, $|\phi^{(2)}_{12}|=1$, $\lambda^{(1)}_1=1$,
$\lambda^{(1)}_2=0$, so that $S_{\rm ent}=0$;
\end{itemize}

{\it To summarize}: the {\it entanglement entropy of the composite fermion} is
 either constant $S_{\rm ent}=\ln 2$ or $S_{\rm ent}=0$ in some special cases, or it is
 given by a general parameter-dependent expression, see~(\ref{S_ent0}).
Let us also remark on the effect of the deformation parameter,
say through $\chi(2)$. Though for each of the considered cases it has not entered
the resp. Schmidt coefficients and entanglement entropy, it can manifest
itself when calculating the averages of physical quantities over quantum states.

\section{More general situations for composite fermions built from fermion and
deformed boson} \label{sec:more_gen}

Before we proceed further
examples generalizing the above ones, let us
make some general remark. Denoting by $D_{CF}$ and $D_f$ the number of modes
respectively for composite fermions and the constituent fermions, we have:
$D_{CF}\le D_f$. Indeed, let $(\alpha_1,...,\alpha_{D_{CF}})$ be the set of all
(differing) CF modes, and let $D_{CF}>D_f$.
Now evaluate the state
\begin{equation}\label{D_QF-prod}
A_{\alpha_1} A_{\alpha_1}^\dag A_{\alpha_2}^\dag ... A_{\alpha_{D_{CF}}}^\dag |0\rangle
= \sum_{\mu\nu...\mu_{D_{CF}}\nu_{D_{CF}}} \overline{\Phi_{\alpha_1}^{\mu\nu}}
\Phi_{\alpha_1}^{\mu\nu}...\Phi_{\alpha_{D_{CF}}}^{\mu_{D_{CF}}\nu_{D_{CF}}}
a_\mu b_\nu a^\dag_{\mu_1}b^\dag_{\nu_1}... a^\dag_{\mu_{D_{CF}}}b^\dag_{\nu_{D_{CF}}}|0\rangle =0.
\end{equation}
Since among $b^\dag_{\nu_1}$,...,$b^\dag_{\nu_{D_{CF}}}$, for
$D_{CF}>D_f$, there are at least two coinciding fermionic creation
operators, that results in zero. 
On the other hand, using realization conditions~(\ref{req2'})-(\ref{req3}) we have
\[
A_{\alpha_1} A_{\alpha_1}^\dag A_{\alpha_2}^\dag ... A_{\alpha_{D_{CF}}}^\dag |0\rangle
= A_{\alpha_2}^\dag ... A_{\alpha_{D_{CF}}}^\dag |0\rangle \ne 0.
\]
The latter inequality holds due to the orthonormality and mode-independence conditions
for (deformed) fermions which realize CFs, see~(\ref{orthonorm}), (\ref{req2}),
(\ref{indep(strict)}). But that contradicts~(\ref{D_QF-prod}). So we conclude that
$D_{CF}\le D_f$. Then, as further directions of the extension of the considered
$D_{CF}=D_f=D_b=2$ case where $D_b$ is the number of modes for the (non-deformed or deformed) constituent boson,
such cases that $D_{CF}=2$, $D_f=D_b=3$, and $D_{CF}=3$, $D_f=3$
can also be treated.

\paragraph{Composite fermions in two modes, with {\it non-deformed} constituents in three modes.}

In this case we take $\alpha=\overline{1,2}$, $\mu,\nu=\overline{1,3}$ so that
$\Phi_\alpha^{\mu\nu}$ are some two $3\times3$-matrices. The realization conditions retain
the form~(\ref{eq1'nd})-(\ref{eq3'nd}). Writing $D_1$ and $\tilde{\Phi}_2$ explicitly as
\begin{equation}\label{repl_3mod}
D_1 = \diag\{\lambda^{(1)}_1,\lambda^{(1)}_2,\lambda^{(1)}_3\},
\quad \tilde{\Phi}_2 = \left(
\begin{array}{ccc}
    \phi_{11} & \phi_{12} & \phi_{13}\\
    \phi_{21} & \phi_{22} & \phi_{23}\\
    \phi_{31} & \phi_{32} & \phi_{33}
\end{array}\right),
\end{equation}
eq.~(\ref{eq2'nd}) yields the system
\begin{equation}\label{eq2'''nd}
((\lambda^{(1)}_i)^2-(\lambda^{(1)}_j)^2)
\phi_{ij} = 0,\quad i\ne j, \ \ 1\le i,j\le 3.
\end{equation}
If the diagonal elements of $D_1$ are different, $\lambda^{(1)}_i\ne \lambda^{(1)}_j$,
$i\ne j$, then $\phi_{ij} = 0$ for $i\ne j$, i.e. matrix $\tilde{\Phi}_2$ is
diagonal too: $\tilde{\Phi}_2 = \diag\{\phi_{11},\phi_{22},\phi_{33}\}$,
$|\phi_{11}|^2+|\phi_{22}|^2+|\phi_{33}|^2 = 1$. The only remaining nontrivial
realization condition is the orthogonality condition in~(\ref{eq1'nd}), which
reduces to the orthogonality condition for the vectors $(\lambda^{(1)}_1,
\lambda^{(1)}_2,\lambda^{(1)}_3)$ and $(\phi_{11},\phi_{22},\phi_{33})$, i.e.
\begin{equation}\label{orth3d-nd}
\lambda^{(1)}_1\overline{\phi_{11}} + \lambda^{(1)}_2\overline{\phi_{22}}
+ \lambda^{(1)}_3\overline{\phi_{33}}=0.
\end{equation}
For the {\it entanglement entropy within a CF} belonging to
each of the two modes we have
\begin{equation}\label{S^(1,2)}
S_{\rm ent}^{(1)}\equiv S_{\rm ent}|_{\alpha=1} = -\sum_{i=1}^3 (\lambda^{(1)}_i)^2
\ln (\lambda^{(1)}_i)^2,\quad
S_{\rm ent}^{(2)}\equiv S_{\rm ent}|_{\alpha=2} = -\sum_{i=1}^3 |\phi_{ii}|^2 \ln |\phi_{ii}|^2.
\end{equation}
It can be parameterized by the angles e.g. in the form
\begin{align}
&\lambda^{(1)}_1 = \cos\theta^{(1)}_1 \cos\theta^{(1)}_2,\quad
\lambda^{(1)}_2 = \cos\theta_1^{(1)} \sin\theta_2^{(1)},\quad
\lambda^{(1)}_3 = \sin\theta_1^{(1)},\label{d_param}\\
&|\phi_{11}| = \cos\theta_1^{(2)} \cos\theta_2^{(2)},\quad |\phi_{22}| = \cos\theta_1^{(2)} \sin\theta_2^{(2)},
\quad |\phi_{33}| = \sin\theta_1^{(2)}.\label{phi_param}
\end{align}
Then the condition~(\ref{orth3d-nd}) gives the following relation between the angles:
\begin{equation}\label{theta2(theta1)}
\cos^2\theta_1^{(2)} = \frac{\sin^2\theta_1^{(1)}}{1-\sin^2\Omega \cos^2\theta_1^{(1)}}
\end{equation}
where the angle $\Omega$ is defined as $\cos 2\Omega \equiv \cos 2\theta_2^{(1)} \cos 2\theta_2^{(2)}
+ \sin 2\theta_2^{(1)} \sin 2\theta_2^{(2)} \cos \gamma'$ with
$\gamma'\equiv \arg (d_1\overline{d_2} \overline{\phi_{11}}\phi_{22})$, and
belonging to the interval $|\theta_2^{(1)}-\theta_2^{(2)}|<\Omega<\theta_2^{(1)}+\theta_2^{(2)}$.
Substituting~(\ref{d_param}) and~(\ref{phi_param}) in~(\ref{S^(1,2)}) and
using (\ref{theta2(theta1)}), we obtain
\begin{align}
&S_{\rm ent}^{(1)} = S_2(\theta_1^{(1)}) + \cos^2\theta_1^{(1)} S_2(\theta_2^{(1)}),\\
&S_{\rm ent}^{(2)} = \frac{S_2(\theta_2^{(2)}) - \ctg^2 \theta_1^{(1)} \cos^2\Omega\,
\ln(\ctg^2 \theta_1^{(1)} \cos^2\Omega)}{1+\ctg^2 \theta_1^{(1)} \cos^2\Omega}
+ \ln(1+\ctg^2 \theta_1^{(1)} \cos^2\Omega)
\end{align}
where the function $S_2(x)$ is defined in~(\ref{S_2-def}).
\par\noindent{\bf Remark}. Another parametrization of two orthonormal vectors
$(\lambda^{(1)}_1,\lambda^{(1)}_2,\lambda^{(1)}_3)$
and $(\phi_{11},\phi_{22},\phi_{33})$ follows from the parametrization of $SU(3)$
since the rows/colums of matrices from $SU(3)$ constitute orthonormal vectors.
Indeed, using the parametrization given in~\cite{Bronzan1988ParamSU3} and retaining
the parametrization~(\ref{d_param}) for $\alpha=1$ mode we have the following
parametrization for the $\alpha=2$ mode ($\theta_{1,2}\equiv \theta_{1,2}^{(1)}$):
\begin{align}
&\phi_{11} = -\sin\theta_1 \cos\theta_2 \cos\theta_3 - \sin\theta_2 \sin\theta_3 e^{i\gamma},\quad
\phi_{22} = \cos\theta_2 \sin\theta_3 e^{i\gamma} - \sin\theta_1 \sin\theta_2 \cos\theta_3,\nonumber\\
&\phi_{33} = \cos\theta_1 \cos\theta_3,\qquad 0\le \theta_1,\theta_2,\theta_3\le \pi/2,\ \
0\le \gamma\le 2\pi.
\end{align}
The corresponding entanglement entropy expressions, $S_{\rm ent}^{(1)}(\theta_1,\theta_2)$
and $S_{\rm ent}^{(2)}(\theta_1,\theta_2,\theta_3,\gamma)$, stem from~(\ref{S^(1,2)}).
To achieve standard ordering $\lambda^{(1)}_1\ge \lambda^{(1)}_2\ge \lambda^{(1)}_3$
we have to impose $0\le \theta_2\le \frac{\pi}{4}$, $0\le \theta_1\le \arctg\sin\theta_2$.

\begin{figure}[h]
\centering
\includegraphics[width=0.49\columnwidth]{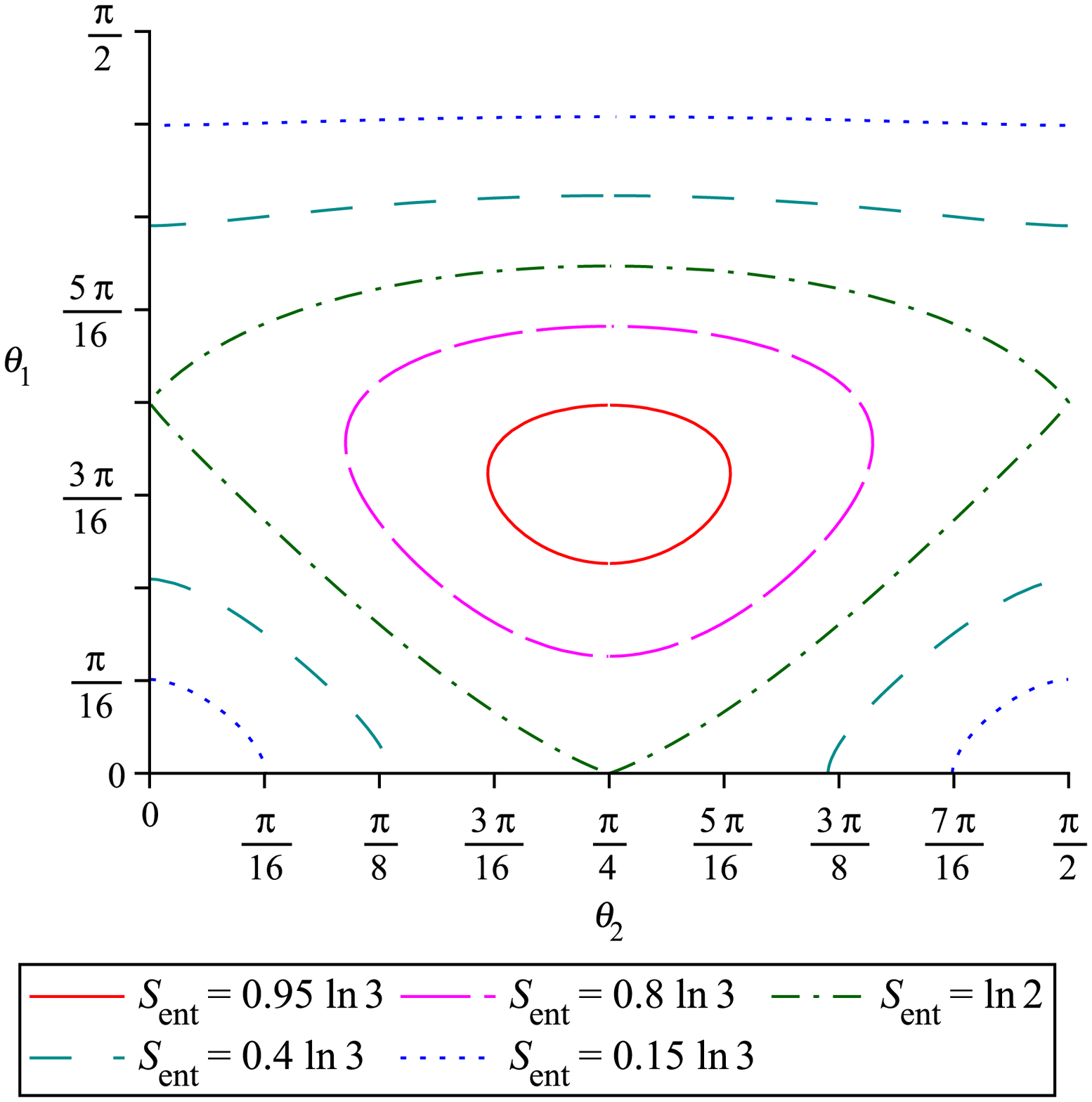}
\includegraphics[width=0.49\columnwidth]{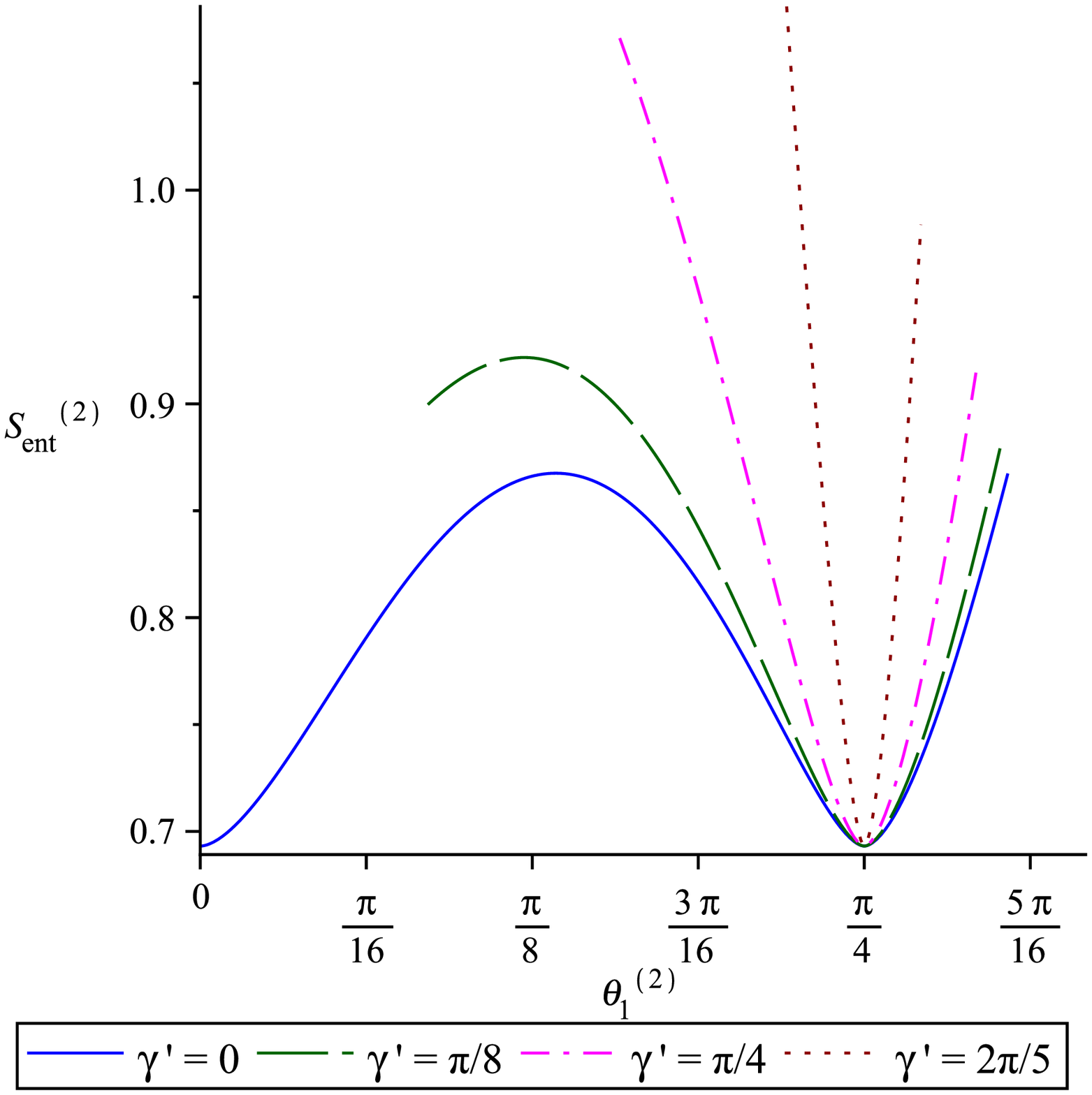}
\caption{Left: Equi-entropic curves (for constant composite
fermion entanglement entropy $S_{\rm ent}^{(\alpha)}$)
versus parameters $\theta_1^{(\alpha)},\theta_2^{(\alpha)}$, at a fixed
mode~$\alpha =1$ or $2$ (the case of three-mode constituents). Right: Entanglement entropy
$S_{\rm ent}^{(2)}(\theta_1^{(2)},\gamma')$ for a composite fermion in $\alpha=2$
mode at fixed entanglement entropy $S_{\rm ent}^{(1)}=\ln 3$ for $\alpha=1$
mode of composite fermion.}
\label{fig3}
\end{figure}

Thus, {\it composite fermion entanglement entropies}~$S_{\rm ent}^{(1)}$, $S_{\rm ent}^{(2)}$
in eq.~(\ref{S^(1,2)}) are parameterized by four angles. Unlike the two-mode $\mu,\nu=\overline{1,2}$
case considered in~Section~\ref{sec:2mod(nd)} where $S_{\rm ent}^{(1)}-S_{\rm ent}^{(2)}=0$,
and $0\le S_{\rm ent}^{(\alpha)}\le \ln 2$, $\alpha=\overline{1,2}$, now it can be shown
for the $\mu,\nu=\overline{1,3}$ case that $|S_{\rm ent}^{(1)}-S_{\rm ent}^{(2)}|\le \ln 2$,
and $0\le S_{\rm ent}^{(\alpha)}\le \ln 3$. This restriction on the difference
$|S_{\rm ent}^{(1)}-S_{\rm ent}^{(2)}|$ can be viewed
as the necessary condition for the realization.
 For the illustration of the dependence $S_{\rm ent}^{(\alpha)} =
S_{\rm ent}^{(\alpha)}(\theta_1^{(\alpha)},\theta_2^{(\alpha)})$ at a fixed mode
$\alpha$ with the other one ignored, equi-entropic curves in the resp.
$\theta_1$-, $\theta_2$-angles are given in Fig.~\ref{fig3} (left). A similar behavior
can be seen e.g. in~\cite{Bolukbasi2006qutrits}, in the context of
the $SU(3)$ parametrization of qutrits.

Let us consider
the case when two diagonal elements of $D_1$, e.g. $\lambda^{(1)}_1$ and
$\lambda^{(1)}_2$ coincide, but differ from the remaining one: $\lambda^{(1)}_1 = \lambda^{(1)}_2
\ne \lambda^{(1)}_3$.
Condition~(\ref{eq2'''nd}) yields $\phi_{13}=\phi_{23}=\phi_{31}=\phi_{32}=0$.
Next, we present the $2\times2$ block $\Bigl({\scriptstyle\phi_{11}\ \phi_{12}\atop\scriptstyle \phi_{21}\ \phi_{22}}\Bigr)$
of $\tilde{\Phi}_2$ using singular value decomposition
applied for $SU(2)$-matrices,
\begin{equation}\label{block}
\left(
\begin{array}{cc}
    \phi_{11} & \phi_{12}\\
    \phi_{21} & \phi_{22}
\end{array}\right)
= e^{i\eta} \tilde{U} \tilde{D}_2 \tilde{V}^\dag = e^{i\eta} \left(
\begin{array}{cc}
    \tilde{u}_1 & \tilde{u}_2\\
    -\overline{\tilde{u}_2} & \overline{\tilde{u}_1}
\end{array}\right) \left(
\begin{array}{cc}
    \lambda^{(2)}_1 & 0\\
    0 & \lambda^{(2)}_2
\end{array}\right) \left(
\begin{array}{cc}
    \tilde{v}_1 & \tilde{v}_2\\
    -\overline{\tilde{v}_2} & \overline{\tilde{v}_1}
\end{array}\right)^\dag
\end{equation}
where the three matrices $\tilde{U}$, $\tilde{D}_2$, and $\tilde{V}$ are shown explicitly.
Then (\ref{eq3'nd}) (at $\lambda^{(1)}_1=\lambda^{(1)}_2\ne0$)
reduces to the equations as in~(\ref{eq3''nd}).
The orthogonality condition in~(\ref{eq1'nd}) then yields
\begin{equation*}
\lambda^{(1)}_1  \lambda^{(2)}_1 (\tilde{u}_1\overline{\tilde{v}_1}
+ \overline{\tilde{u}_2}\tilde{v}_2) + \lambda^{(1)}_1
\lambda^{(2)}_2 (\overline{\tilde{u}_1}\tilde{v}_1
+ \tilde{u}_2\overline{\tilde{v}_2})
= -e^{-i\eta} \lambda^{(1)}_3 \phi_{33}.
\end{equation*}

\noindent$\bullet$ \ If $\lambda^{(2)}_1=\lambda^{(2)}_2$,
eqs. from~(\ref{eq3''nd}) are satisfied
while the block~(\ref{block}) is proportional to a unitary matrix,
\begin{equation*}
\left(
\begin{array}{cc}
    \phi_{11} & \phi_{12}\\
    \phi_{21} & \phi_{22}
\end{array}\right)\biggr|_{\lambda^{(2)}_1=\lambda^{(2)}_2}
= \lambda^{(2)}_1 e^{i\eta} U',\quad
U' = \left(
\begin{array}{cc}
    u'_1 & u'_2\\
    -\overline{u'_2} & \overline{u'_1}
\end{array}\right) \in SU(2),
\end{equation*}
that leads to the orthogonality condition
\begin{equation*}
\lambda^{(1)}_1 \lambda^{(2)}_1 e^{i\eta}
(u'_1+ \overline{u'_1})
\!=\! - \lambda^{(1)}_3 \phi_{33} \ \Rightarrow\
2 \lambda^{(2)}_1 \lambda^{(1)}_1 |{\rm Re}\, u'_1|
\!=\! \sqrt{1\!-\!2 (\lambda^{(1)}_1)^2} \sqrt{1\!-\!2 (\lambda^{(2)}_1)^2}.
\end{equation*}
The solution for $\Phi_1$ and $\Phi_2$ is then written as
\begin{equation}
\begin{aligned}
&\Phi_1 = U_1 \diag\{\lambda^{(1)}_1,\lambda^{(1)}_2,
\lambda^{(1)}_3\} V_1 = \tilde{U}_1 \left(
\begin{array}{cc}
    \lambda^{(1)}_1 e^{i\gamma_d} (U'')^\dag & 0\\
    0 & \lambda^{(1)}_3
\end{array}\right) V_1,\\
&\Phi_2 = U_1 \left(
\begin{array}{cc}
    \lambda^{(2)}_1 e^{i\eta} U' & 0\\
    0 & \phi_{33}
\end{array}\right) V_1 = \tilde{U}_1 \left(
\begin{array}{cc}
    \lambda^{(2)}_1 e^{i\eta} U'' & 0\\
    0 & \phi_{33}
\end{array}\right) V_1,
\end{aligned}
\end{equation}
where
\begin{equation*}
\tilde{U}_1 = U_1 \left(
\begin{array}{cc}
    U_d U'' & 0\\
    0 & 1
\end{array}\right),\ \ \
\left(
\begin{array}{cc}
    \lambda^{(1)}_1 & 0\\
    0 & \lambda^{(1)}_2
\end{array}\right) \equiv \lambda^{(1)}_1 e^{i\gamma_d} U_d,
\ \ \ U_d^\dag U' = (U'')^2,\ \ U_d, U''\in SU(2).
\end{equation*}
For the {\it entanglement entropy of composite fermion} in this subcase we find
\begin{align}
&S_{\rm ent}^{(1)} = -\cos^2\theta_1 \ln\Bigl(\frac12 \cos^2\theta_1\Bigr)
- \sin^2\theta_1 \ln \sin^2\theta_1 = \cos^2\theta_1 \ln 2 + S_2(\theta_1),\label{S^(1)_2eq}\\
&S_{\rm ent}^{(2)} = \ln\Bigl(\tg^2\theta_1 \!+\! \frac14|\Tr U'|^2\Bigr)
- \frac{\tg^2\theta_1 \ln\bigl(\frac12 \tg^2\theta_1\bigr) \!+\! \frac14|\Tr U'|^2
\ln\bigl(\frac14|\Tr U'|^2\bigr)}{\tg^2\theta_1 + \frac14|\Tr U'|^2},\ \ \ 0\le |\Tr U'|\le 2.
\end{align}

\noindent$\bullet$ \ If $\lambda^{(2)}_1\ne \lambda^{(2)}_2$,
from eqs.~in~(\ref{eq3''nd}) we obtain
\begin{equation*}
\lambda^{(1)}_1(|u_1|-|v_1|)=0;\quad \lambda^{(1)}_1 (
\tilde{u}_1 \tilde{u}_2 - \tilde{v}_1 \tilde{v}_2) = 0\ \
\mathop{\Rightarrow}\limits_{\lambda^{(1)}_1\ne0}\ \ \lambda^{(1)}_1
\lambda^{(2)}_1 \frac{\tilde{u}_1}{\tilde{v}_1}
+ \lambda^{(1)}_1 \lambda^{(2)}_2
\frac{\overline{\tilde{u}_1}}{\overline{\tilde{v}_1}}
= -e^{-i\eta} \lambda^{(1)}_3 \phi_{33}.
\end{equation*}
Let $\frac{\tilde{u}_1}{\tilde{v}_1} = e^{i\delta}$.
Then $\tilde{V} = \tilde{U} \diag\{e^{-i\delta},e^{i\delta}\}$
and the involved parameters are related as
\begin{equation}
\lambda^{(1)}_1|\lambda^{(2)}_1 e^{i\delta}
+ \lambda^{(2)}_2 e^{-i\delta}|
= \sqrt{1\!-\!2 (\lambda^{(1)}_1)^2} \sqrt{1\!-\!(\lambda^{(2)}_1)^2\!-\!(\lambda^{(2)}_2)^2}.
\end{equation}
The {\it corresponding expression for the entanglement entropy} $S_{\rm ent}^{(2)}$
for the $\alpha=2$ mode reads
\begin{multline}\label{S^(2)_2eq'}
S_{\rm ent}^{(2)} = \ln 2 - |\sin\theta_1\cos\theta_3 + \sin\theta_3 e^{i\gamma}|^2
\ln |\sin\theta_1\cos\theta_3 + \sin\theta_3 e^{i\gamma}| -\\
- |\sin\theta_1\cos\theta_3 - \sin\theta_3 e^{i\gamma}|^2 \ln |\sin\theta_1\cos\theta_3 - \sin\theta_3 e^{i\gamma}|
- \cos^2\theta_1\cos^2\theta_3 \ln (2 \cos^2\theta_1\cos^2\theta_3),
\end{multline}
while $S_{\rm ent}^{(1)}$ is given in~(\ref{S^(1)_2eq}).

For the case $\lambda^{(1)}_1 = \lambda^{(1)}_2 = \lambda^{(1)}_3$
matrix $\tilde{\Phi}_2$ satisfies relations~(\ref{norm_sol}). Presenting
$\tilde{\Phi}_2$ as in~(\ref{Phi2tilde-def}) with $D_2=\diag\{\lambda^{(2)}_1,\lambda^{(2)}_2,\lambda^{(2)}_3\}$,
$\tilde{U},\tilde{V}\in SU(3)$, we obtain the equations similar to~(\ref{eq1'nd}), (\ref{eq2'nd}):
\begin{equation}\label{eq_D2}
D_2^2 W = W D_2^2,\quad \Tr (D_2 W)=0,\quad \Tr D_2^2=1,\quad
W= \tilde{V}^\dag \tilde{U}\in SU(3).
\end{equation}
If $\lambda^{(2)}_i\ne \lambda^{(2)}_j$, $i\ne j$, from equation analogous
to~(\ref{eq2'''nd}) we have the solution
\begin{equation}
W = \diag\{w_{11},w_{22},w_{33}\},\quad |w_{ii}|=1,\ \ i=\overline{1,3},
\quad \sum\nolimits_i \lambda^{(2)}_i w_{ii} =0,
\end{equation}
so that $\tilde{\Phi}_2 = e^{i\eta} \tilde{U}\, D_2 W\, \tilde{U}^\dag$.
If $\lambda^{(2)}_1\!=\! \lambda^{(2)}_2\!\ne\! \lambda^{(2)}_3$, matrix $W$ is
block-diagonal, $W = \diag\{w_{33}^{-1/2} W',w_{33}\}$, $|w_{33}|=1$, $W'\in SU(2)$.
From the second equation in~(\ref{eq_D2}) we have
\begin{equation*}
\lambda^{(2)}_1 w_{33}^{-1/2} \Tr W' + \lambda^{(2)}_3 w_{33} = 0 \quad
\Rightarrow\quad \lambda^{(2)}_3 = |\Tr W'| \lambda^{(2)}_1,\ \ \ 0\le |\Tr W'|\le 2,
\end{equation*}
so that $\tilde{\Phi}_2 = e^{i\eta} \tilde{U} \diag\{w_{33}^{-1/2} \lambda^{(2)}_1 W',
w_{33} \lambda^{(2)}_3\} \tilde{U}^\dag$.
If $\lambda^{(2)}_1\!=\! \lambda^{(2)}_2\!=\! \lambda^{(2)}_3$: $\tilde{\Phi}_2 = e^{i\eta}
\lambda^{(2)}_1 U'$, $U'\in SU(3)$, $\Tr U'=0$.
 The entanglement entropy for
a CF in $\alpha=1$ mode is $S_{\rm ent}|_{\alpha=1} = \ln 3$.
The {\it entanglement entropy} within a CF in the $\alpha=2$ mode,
for the particular diagonal solution $\tilde{\Phi}_2$ reads
\begin{equation}\label{S^(2)_3eq}
S_{\rm ent}^{(2)} = S_2(\theta_1^{(2)}) + \cos^2\theta_1^{(2)} \Bigl(
\sqrt{1-4K^2} \ln\frac{2|K|}{1+\sqrt{1-4K^2}} - \ln|K|\Bigr),
\ \ \ K=\frac{\sin^2\theta_1^{(2)}-1/2}{\cos^2\theta_1^{(2)} \cos\gamma'},
\end{equation}
so, it takes its values from the interval
$[\ln 2,\,\ln 3]$, see~Fig.~\ref{fig3} (right). Note that expression~(\ref{S^(2)_3eq})
corresponds to~(\ref{S^(2)_2eq'}) at $\sin\theta_1= \frac{1}{\sqrt{3}}$, which for
$\gamma=0$ takes simple symmetric form
\begin{equation}
S_{\rm ent}^{(2)} = \tilde{s}(\theta_3) + \tilde{s}\Bigl(\theta_3+\frac{2\pi}{3}\Bigr)
+ \tilde{s}\Bigl(\theta_3-\frac{2\pi}{3}\Bigr),\quad
\tilde{s}(\theta) \equiv -\frac23 \cos^2\theta \ln \Bigl(\frac23 \cos^2\theta\Bigr).
\end{equation}
 For the particular solution $\tilde{\Phi}_2$
with two equal singular values (or Schmidt coefficients) different from the third one we find
\begin{equation}
S_{\rm ent}^{(2)} = \ln (2+|\Tr W'|^2) - \frac{|\Tr W'|^2}{2+|\Tr W'|^2} \ln |\Tr W'|^2,
\quad 0\le |\Tr W'|\le 2,
\end{equation}
and $S_{\rm ent}^{(2)}$ belongs to interval $[\ln 2,\ln 3]$.
For equal coefficients $\lambda^{(2)}_i$  we have $S_{\rm ent}^{(2)} = \ln 3$.

\section{Discussion and outlook} \label{sec:discussion}

 Let us make few comments on the above results.
  After the problem of realization of composite fermions (CFs) by usual
  fermions was settled, we have explored the topic of main interest
  in this paper: the bipartite entanglement (within the CF)
  measured by the entanglement entropy of CF.
   We have performed our analysis in the two relatively simple cases:
   of one-mode and of two-mode CFs.
   Already the latter case turns out to be nontrivial, implying
   a number of subcases.

    In the entanglement entropy of CFs of the type
    ``fermion + deformed boson'' the very constituent boson deformation does not manifest itself
    explicitly in these one- and two-mode cases, contrary to the earlier studied
    (entanglement entropy of) quasibosons where in the focus was just
    the dependence on deformation parameter~$f$.
    Nevertheless, in the present case there are the parameters
    being involved in the matrix $\Phi$ of the ansatz (\ref{ansatz}), which the entanglement entropy
    of CFs depends upon. This dependence is shown in Fig.~\ref{fig2}. Also
    noteworthy are the properties of CF entanglement entropy pictured in Fig.~\ref{fig3}.

Let us note once more that the results of this paper give
explicit formulas, or constant values in a few cases, for the
entanglement entropy of {\it individual composite fermion}
 (i.e. for the entanglement between constituents), see also Introduction.
 In contrast, the authors of~\cite{Gioev2006Entanglement,Shao2014Entanglement}
 explored entanglement entropy of many-fermion systems in certain space region.
 For instance, in~\cite{Shao2014Entanglement} an efficient numerical methods (improved
Monte-Carlo) were applied to the system of 37 composite fermions,
and the linear size of subsystem entered final result for the
entanglement entropy.

What was the role of deformation parameter~$f$ in the situation with
quasi-bosons? Therein~\cite{GM_Entang,GM_Ent(En)}, we had quite natural feature: the
entanglement entropy was rising with decreasing values of~$f$, i.e.
with the approaching to truly bosonic behavior, either for the Fock
states at fixed mode or for the coherent states.
 In the present case of CFs, we have not yet established possible physical
 meaning of the parameter(s) which the entanglement entropy (and purity)
 depends on, and that of course remains to be done.
Besides, what concerns the important dependence of the entanglement entropy
of CFs on their energy to be yet obtained, such dependence may
have interesting physical consequences
including comparison with the case of quasi-bosons (studied in~\cite{GM_Ent(En)}).
 We hope to obtain such a relation along with its implications in the sequel.

Concerning some experimental testing
of the obtained results we can only mention possible application of
these results to a description of relevant properties of such systems
as ``exciton + electron'' or ``exciton + hole''. Also, there may be a useful
impact on the baryons when these are viewed as diquark-quark systems~\cite{Anselmino_Diquarks}.
At last, let us note that it is also of interest to study another
CF system, that is the composite one
of the type ``fermion + fermion + fermion'', and we intend to report
on that in a near future.

\section*{Acknowledgements}

The research was partially supported by the Special Program of the Division of
Physics and Astronomy of NAS of Ukraine.

\bibliography{references}

\providecommand{\noopsort}[1]{}\providecommand{\singleletter}[1]{#1}%
\begin{thebibliography}{29}%
\makeatletter
\providecommand \@ifxundefined [1]{%
 \@ifx{#1\undefined}
}%
\providecommand \@ifnum [1]{%
 \ifnum #1\expandafter \@firstoftwo
 \else \expandafter \@secondoftwo
 \fi
}%
\providecommand \@ifx [1]{%
 \ifx #1\expandafter \@firstoftwo
 \else \expandafter \@secondoftwo
 \fi
}%
\providecommand \natexlab [1]{#1}%
\providecommand \enquote  [1]{``#1''}%
\providecommand \bibnamefont  [1]{#1}%
\providecommand \bibfnamefont [1]{#1}%
\providecommand \citenamefont [1]{#1}%
\providecommand \href@noop [0]{\@secondoftwo}%
\providecommand \href [0]{\begingroup \@sanitize@url \@href}%
\providecommand \@href[1]{\@@startlink{#1}\@@href}%
\providecommand \@@href[1]{\endgroup#1\@@endlink}%
\providecommand \@sanitize@url [0]{\catcode `\\12\catcode `\$12\catcode
  `\&12\catcode `\#12\catcode `\^12\catcode `\_12\catcode `\%12\relax}%
\providecommand \@@startlink[1]{}%
\providecommand \@@endlink[0]{}%
\providecommand \url  [0]{\begingroup\@sanitize@url \@url }%
\providecommand \@url [1]{\endgroup\@href {#1}{\urlprefix }}%
\providecommand \urlprefix  [0]{URL }%
\providecommand \Eprint [0]{\href }%
\providecommand \doibase [0]{http://dx.doi.org/}%
\providecommand \selectlanguage [0]{\@gobble}%
\providecommand \bibinfo  [0]{\@secondoftwo}%
\providecommand \bibfield  [0]{\@secondoftwo}%
\providecommand \translation [1]{[#1]}%
\providecommand \BibitemOpen [0]{}%
\providecommand \bibitemStop [0]{}%
\providecommand \bibitemNoStop [0]{.\EOS\space}%
\providecommand \EOS [0]{\spacefactor3000\relax}%
\providecommand \BibitemShut  [1]{\csname bibitem#1\endcsname}%
\let\auto@bib@innerbib\@empty
\bibitem [{\citenamefont {Jain}(2007)}]{Jain2007Composite}%
  \BibitemOpen
  \bibfield  {author} {\bibinfo {author} {\bibfnamefont {J.~K.}\ \bibnamefont
  {Jain}},\ }\href@noop {} {\textit {\bibinfo {title} {Composite fermions}}}\
  (\bibinfo  {publisher} {Cambridge: Cambridge University Press},\ \bibinfo
  {year} {2007})\BibitemShut {NoStop}%
\bibitem [{\citenamefont {Hadjimichef}\ \textit {et~al.}(1998)\citenamefont
  {Hadjimichef}, \citenamefont {Krein}, \citenamefont {Szpigel},\ and\
  \citenamefont {Veiga}}]{Hadjimichef1998}%
  \BibitemOpen
  \bibfield  {author} {\bibinfo {author} {\bibfnamefont {D.}~\bibnamefont
  {Hadjimichef}}, \bibinfo {author} {\bibfnamefont {G.}~\bibnamefont {Krein}},
  \bibinfo {author} {\bibfnamefont {S.}~\bibnamefont {Szpigel}}, \ and\
  \bibinfo {author} {\bibfnamefont {J.~D.}\ \bibnamefont {Veiga}},\ }\href
  {\doibase 10.1006/aphy.1998.5825} {\bibfield  {journal} {\bibinfo  {journal}
  {Ann. Phys.}\ }\textbf {\bibinfo {volume} {268}},\ \bibinfo {pages} {105}
  (\bibinfo {year} {1998})}\BibitemShut {NoStop}%
\bibitem [{\citenamefont {Oh}\ and\ \citenamefont {Kim}(2004)}]{Oh2004Penta}%
  \BibitemOpen
  \bibfield  {author} {\bibinfo {author} {\bibfnamefont {Y.}~\bibnamefont
  {Oh}}\ and\ \bibinfo {author} {\bibfnamefont {H.}~\bibnamefont {Kim}},\
  }\href {\doibase 10.1103/PhysRevD.70.094022} {\bibfield  {journal} {\bibinfo
  {journal} {Phys. Rev. D}\ }\textbf {\bibinfo {volume} {70}},\ \bibinfo
  {pages} {094022} (\bibinfo {year} {2004})}\BibitemShut {NoStop}%
\bibitem [{\citenamefont {Browder}\ \textit {et~al.}(2004)\citenamefont
  {Browder}, \citenamefont {Klebanov},\ and\ \citenamefont
  {Marlow}}]{Browder2004Penta}%
  \BibitemOpen
  \bibfield  {author} {\bibinfo {author} {\bibfnamefont {T.~E.}\ \bibnamefont
  {Browder}}, \bibinfo {author} {\bibfnamefont {I.~R.}\ \bibnamefont
  {Klebanov}}, \ and\ \bibinfo {author} {\bibfnamefont {D.~R.}\ \bibnamefont
  {Marlow}},\ }\href {\doibase
  http://dx.doi.org/10.1016/j.physletb.2004.03.003} {\bibfield  {journal}
  {\bibinfo  {journal} {Phys. Lett. B}\ }\textbf {\bibinfo {volume} {587}},\
  \bibinfo {pages} {62 } (\bibinfo {year} {2004})}\BibitemShut {NoStop}%
\bibitem [{\citenamefont {Tichy}\ \textit {et~al.}(2011)\citenamefont {Tichy},
  \citenamefont {Mintert},\ and\ \citenamefont {Buchleitner}}]{Tichy_Rev}%
  \BibitemOpen
  \bibfield  {author} {\bibinfo {author} {\bibfnamefont {M.~C.}\ \bibnamefont
  {Tichy}}, \bibinfo {author} {\bibfnamefont {F.}~\bibnamefont {Mintert}}, \
  and\ \bibinfo {author} {\bibfnamefont {A.}~\bibnamefont {Buchleitner}},\
  }\href {http://stacks.iop.org/0953-4075/44/i=19/a=192001} {\bibfield
  {journal} {\bibinfo  {journal} {J. Phys. B: At. Mol. Opt. Phys.}\ }\textbf
  {\bibinfo {volume} {44}},\ \bibinfo {pages} {192001} (\bibinfo {year}
  {2011})}\BibitemShut {NoStop}%
\bibitem [{\citenamefont {Avancini}\ and\ \citenamefont {Krein}(1995)}]{Avan}%
  \BibitemOpen
  \bibfield  {author} {\bibinfo {author} {\bibfnamefont {S.~S.}\ \bibnamefont
  {Avancini}}\ and\ \bibinfo {author} {\bibfnamefont {G.}~\bibnamefont
  {Krein}},\ }\href {\doibase 10.1088/0305-4470/28/3/021} {\bibfield  {journal}
  {\bibinfo  {journal} {J. Phys. A: Math. Gen.}\ }\textbf {\bibinfo {volume}
  {28}},\ \bibinfo {pages} {685} (\bibinfo {year} {1995})}\BibitemShut
  {NoStop}%
\bibitem [{\citenamefont {Perkins}(2002)}]{Perkins_2002}%
  \BibitemOpen
  \bibfield  {author} {\bibinfo {author} {\bibfnamefont {W.~A.}\ \bibnamefont
  {Perkins}},\ }\href {http://dx.doi.org/10.1023/A:1015728722664} {\bibfield
  {journal} {\bibinfo  {journal} {Int. J. Theor. Phys.}\ }\textbf {\bibinfo
  {volume} {41}},\ \bibinfo {pages} {823} (\bibinfo {year} {2002})}\BibitemShut
  {NoStop}%
\bibitem [{\citenamefont {Moskalenko}\ and\ \citenamefont
  {Snoke}(2000)}]{Moskalenko2000Bose}%
  \BibitemOpen
  \bibfield  {author} {\bibinfo {author} {\bibfnamefont {S.~A.}\ \bibnamefont
  {Moskalenko}}\ and\ \bibinfo {author} {\bibfnamefont {D.~W.}\ \bibnamefont
  {Snoke}},\ }\href@noop {} {\textit {\bibinfo {title} {Bose-Einstein
  condensation of excitons and biexcitons: and coherent nonlinear optics with
  excitons}}}\ (\bibinfo  {publisher} {Cambridge Univ. Press},\ \bibinfo
  {address} {Cambridge, UK},\ \bibinfo {year} {2000})\BibitemShut {NoStop}%
\bibitem [{\citenamefont {Bethe}\ and\ \citenamefont
  {Salpeter}(1957)}]{Bethe_Salpeter}%
  \BibitemOpen
  \bibfield  {author} {\bibinfo {author} {\bibfnamefont {H.~A.}\ \bibnamefont
  {Bethe}}\ and\ \bibinfo {author} {\bibfnamefont {E.~E.}\ \bibnamefont
  {Salpeter}},\ }\href@noop {} {\textit {\bibinfo {title} {Quantum Mechanics of
  One- and Two-Electron Atoms}}}\ (\bibinfo  {publisher} {Springer-Verlag},\
  \bibinfo {address} {Berlin},\ \bibinfo {year} {1957})\BibitemShut {NoStop}%
\bibitem [{\citenamefont {Esquivel}\ \textit {et~al.}(2011)\citenamefont
  {Esquivel}, \citenamefont {Flores-Gallegos}, \citenamefont {Molina-Espiritu},
  \citenamefont {Plastino}, \citenamefont {Angulo}, \citenamefont {Antolin},\
  and\ \citenamefont {Dehesa}}]{Esquivel}%
  \BibitemOpen
  \bibfield  {author} {\bibinfo {author} {\bibfnamefont {R.~O.}\ \bibnamefont
  {Esquivel}}, \bibinfo {author} {\bibfnamefont {N.}~\bibnamefont
  {Flores-Gallegos}}, \bibinfo {author} {\bibfnamefont {M.}~\bibnamefont
  {Molina-Espiritu}}, \bibinfo {author} {\bibfnamefont {A.~R.}\ \bibnamefont
  {Plastino}}, \bibinfo {author} {\bibfnamefont {J.~C.}\ \bibnamefont
  {Angulo}}, \bibinfo {author} {\bibfnamefont {J.}~\bibnamefont {Antolin}}, \
  and\ \bibinfo {author} {\bibfnamefont {J.~S.}\ \bibnamefont {Dehesa}},\
  }\href {http://stacks.iop.org/0953-4075/44/i=17/a=175101} {\bibfield
  {journal} {\bibinfo  {journal} {J. Phys. B: At. Mol. Opt. Phys.}\ }\textbf
  {\bibinfo {volume} {44}},\ \bibinfo {pages} {175101} (\bibinfo {year}
  {2011})}\BibitemShut {NoStop}%
\bibitem [{\citenamefont {Gavrilik}\ and\ \citenamefont
  {Mishchenko}(2012)}]{GM_Entang}%
  \BibitemOpen
  \bibfield  {author} {\bibinfo {author} {\bibfnamefont {A.~M.}\ \bibnamefont
  {Gavrilik}}\ and\ \bibinfo {author} {\bibfnamefont {{\relax Yu}.~A.}\
  \bibnamefont {Mishchenko}},\ }\href {\doibase 10.1016/j.physleta.2012.03.053}
  {\bibfield  {journal} {\bibinfo  {journal} {Phys. Lett. A}\ }\textbf
  {\bibinfo {volume} {376}},\ \bibinfo {pages} {1596 } (\bibinfo {year}
  {2012})}\BibitemShut {NoStop}%
\bibitem [{\citenamefont {Gavrilik}\ and\ \citenamefont
  {Mishchenko}(2013)}]{GM_Ent(En)}%
  \BibitemOpen
  \bibfield  {author} {\bibinfo {author} {\bibfnamefont {A.~M.}\ \bibnamefont
  {Gavrilik}}\ and\ \bibinfo {author} {\bibfnamefont {{\relax Yu}.~A.}\
  \bibnamefont {Mishchenko}},\ }\href
  {http://stacks.iop.org/1751-8121/46/i=14/a=145301} {\bibfield  {journal}
  {\bibinfo  {journal} {J. Phys. A: Math. Theor.}\ }\textbf {\bibinfo {volume}
  {46}},\ \bibinfo {pages} {145301} (\bibinfo {year} {2013})}\BibitemShut
  {NoStop}%
\bibitem [{\citenamefont {Gavrilik}\ \textit
  {et~al.}(2011{\natexlab{a}})\citenamefont {Gavrilik}, \citenamefont
  {Kachurik},\ and\ \citenamefont {Mishchenko}}]{GKM2}%
  \BibitemOpen
  \bibfield  {author} {\bibinfo {author} {\bibfnamefont {A.~M.}\ \bibnamefont
  {Gavrilik}}, \bibinfo {author} {\bibfnamefont {I.~I.}\ \bibnamefont
  {Kachurik}}, \ and\ \bibinfo {author} {\bibfnamefont {{\relax Yu}.~A.}\
  \bibnamefont {Mishchenko}},\ }\href {\doibase 10.1088/1751-8113/44/47/475303}
  {\bibfield  {journal} {\bibinfo  {journal} {J. Phys. A: Math. Theor.}\
  }\textbf {\bibinfo {volume} {44}},\ \bibinfo {pages} {475303} (\bibinfo
  {year} {2011}{\natexlab{a}})}\BibitemShut {NoStop}%
\bibitem [{\citenamefont {Gavrilik}\ \textit
  {et~al.}(2011{\natexlab{b}})\citenamefont {Gavrilik}, \citenamefont
  {Kachurik},\ and\ \citenamefont {Mishchenko}}]{GKM}%
  \BibitemOpen
  \bibfield  {author} {\bibinfo {author} {\bibfnamefont {A.~M.}\ \bibnamefont
  {Gavrilik}}, \bibinfo {author} {\bibfnamefont {I.~I.}\ \bibnamefont
  {Kachurik}}, \ and\ \bibinfo {author} {\bibfnamefont {{\relax Yu}.~A.}\
  \bibnamefont {Mishchenko}},\ }\href
  {http://www.ujp.bitp.kiev.ua/files/file/papers/56/9/560911p.pdf} {\bibfield
  {journal} {\bibinfo  {journal} {Ukr. J. Phys.}\ }\textbf {\bibinfo {volume}
  {56}},\ \bibinfo {pages} {948} (\bibinfo {year}
  {2011}{\natexlab{b}})}\BibitemShut {NoStop}%
\bibitem [{\citenamefont {Gavrilik}\ and\ \citenamefont
  {Mishchenko}(2015)}]{GM2015muq-corr}%
  \BibitemOpen
  \bibfield  {author} {\bibinfo {author} {\bibfnamefont {A.~M.}\ \bibnamefont
  {Gavrilik}}\ and\ \bibinfo {author} {\bibfnamefont {{\relax Yu}.~A.}\
  \bibnamefont {Mishchenko}},\ }\href {\doibase
  http://dx.doi.org/10.1016/j.nuclphysb.2014.12.017} {\bibfield  {journal}
  {\bibinfo  {journal} {Nucl. Phys. B}\ }\textbf {\bibinfo {volume} {891}},\
  \bibinfo {pages} {466 } (\bibinfo {year} {2015})}\BibitemShut {NoStop}%
\bibitem [{\citenamefont {Horodecki}\ \textit {et~al.}(2009)\citenamefont
  {Horodecki}, \citenamefont {Horodecki}, \citenamefont {Horodecki},\ and\
  \citenamefont {Horodecki}}]{Horodecki}%
  \BibitemOpen
  \bibfield  {author} {\bibinfo {author} {\bibfnamefont {R.}~\bibnamefont
  {Horodecki}}, \bibinfo {author} {\bibfnamefont {P.}~\bibnamefont
  {Horodecki}}, \bibinfo {author} {\bibfnamefont {M.}~\bibnamefont
  {Horodecki}}, \ and\ \bibinfo {author} {\bibfnamefont {K.}~\bibnamefont
  {Horodecki}},\ }\href {\doibase 10.1103/RevModPhys.81.865} {\bibfield
  {journal} {\bibinfo  {journal} {Rev. Mod. Phys.}\ }\textbf {\bibinfo {volume}
  {81}},\ \bibinfo {pages} {865 } (\bibinfo {year} {2009})}\BibitemShut
  {NoStop}%
\bibitem [{\citenamefont {Law}(2005)}]{Law}%
  \BibitemOpen
  \bibfield  {author} {\bibinfo {author} {\bibfnamefont {C.~K.}\ \bibnamefont
  {Law}},\ }\href {\doibase 10.1103/PhysRevA.71.034306} {\bibfield  {journal}
  {\bibinfo  {journal} {Phys.\ Rev. A}\ }\textbf {\bibinfo {volume} {71}},\
  \bibinfo {pages} {034306} (\bibinfo {year} {2005})}\BibitemShut {NoStop}%
\bibitem [{\citenamefont {Chudzicki}\ \textit {et~al.}(2010)\citenamefont
  {Chudzicki}, \citenamefont {Oke},\ and\ \citenamefont
  {Wootters}}]{Chudzicki}%
  \BibitemOpen
  \bibfield  {author} {\bibinfo {author} {\bibfnamefont {C.}~\bibnamefont
  {Chudzicki}}, \bibinfo {author} {\bibfnamefont {O.}~\bibnamefont {Oke}}, \
  and\ \bibinfo {author} {\bibfnamefont {W.~K.}\ \bibnamefont {Wootters}},\
  }\href {\doibase 10.1103/PhysRevLett.104.070402} {\bibfield  {journal}
  {\bibinfo  {journal} {Phys. Rev. Lett.}\ }\textbf {\bibinfo {volume} {104}},\
  \bibinfo {pages} {070402} (\bibinfo {year} {2010})}\BibitemShut {NoStop}%
\bibitem [{\citenamefont {Ramanathan}\ \textit {et~al.}(2011)\citenamefont
  {Ramanathan}, \citenamefont {Kurzynski}, \citenamefont {Chuan}, \citenamefont
  {Santos},\ and\ \citenamefont {Kaszlikowski}}]{Ramanathan}%
  \BibitemOpen
  \bibfield  {author} {\bibinfo {author} {\bibfnamefont {R.}~\bibnamefont
  {Ramanathan}}, \bibinfo {author} {\bibfnamefont {P.}~\bibnamefont
  {Kurzynski}}, \bibinfo {author} {\bibfnamefont {T.~K.}\ \bibnamefont
  {Chuan}}, \bibinfo {author} {\bibfnamefont {M.~F.}\ \bibnamefont {Santos}}, \
  and\ \bibinfo {author} {\bibfnamefont {D.}~\bibnamefont {Kaszlikowski}},\
  }\href {\doibase 10.1103/PhysRevA.84.034304} {\bibfield  {journal} {\bibinfo
  {journal} {Phys. Rev. A}\ }\textbf {\bibinfo {volume} {84}},\ \bibinfo
  {pages} {034304} (\bibinfo {year} {2011})}\BibitemShut {NoStop}%
\bibitem [{\citenamefont {Morimae}(2010)}]{Morimae}%
  \BibitemOpen
  \bibfield  {author} {\bibinfo {author} {\bibfnamefont {T.}~\bibnamefont
  {Morimae}},\ }\href {\doibase 10.1103/PhysRevA.81.060304} {\bibfield
  {journal} {\bibinfo  {journal} {Phys. Rev. A}\ }\textbf {\bibinfo {volume}
  {81}},\ \bibinfo {pages} {060304} (\bibinfo {year} {2010})}\BibitemShut
  {NoStop}%
\bibitem [{\citenamefont {Weder}(2011)}]{Weder}%
  \BibitemOpen
  \bibfield  {author} {\bibinfo {author} {\bibfnamefont {R.}~\bibnamefont
  {Weder}},\ }\href {\doibase 10.1103/PhysRevA.84.062320} {\bibfield  {journal}
  {\bibinfo  {journal} {Phys. Rev. A}\ }\textbf {\bibinfo {volume} {84}},\
  \bibinfo {pages} {062320} (\bibinfo {year} {2011})}\BibitemShut {NoStop}%
\bibitem [{\citenamefont {Kurzynski}\ \textit {et~al.}(2012)\citenamefont
  {Kurzynski}, \citenamefont {Ramanathan}, \citenamefont {Soeda}, \citenamefont
  {Chuan},\ and\ \citenamefont {Kaszlikowski}}]{Kurzynski}%
  \BibitemOpen
  \bibfield  {author} {\bibinfo {author} {\bibfnamefont {P.}~\bibnamefont
  {Kurzynski}}, \bibinfo {author} {\bibfnamefont {R.}~\bibnamefont
  {Ramanathan}}, \bibinfo {author} {\bibfnamefont {A.}~\bibnamefont {Soeda}},
  \bibinfo {author} {\bibfnamefont {T.~K.}\ \bibnamefont {Chuan}}, \ and\
  \bibinfo {author} {\bibfnamefont {D.}~\bibnamefont {Kaszlikowski}},\ }\href
  {http://stacks.iop.org/1367-2630/14/i=9/a=093047} {\bibfield  {journal}
  {\bibinfo  {journal} {New J. Phys.}\ }\textbf {\bibinfo {volume} {14}},\
  \bibinfo {pages} {093047} (\bibinfo {year} {2012})}\BibitemShut {NoStop}%
\bibitem [{\citenamefont {Bartley}\ \textit {et~al.}(2013)\citenamefont
  {Bartley}, \citenamefont {Crowley}, \citenamefont {Datta}, \citenamefont
  {Nunn}, \citenamefont {Zhang},\ and\ \citenamefont {Walmsley}}]{Bartley}%
  \BibitemOpen
  \bibfield  {author} {\bibinfo {author} {\bibfnamefont {T.~J.}\ \bibnamefont
  {Bartley}}, \bibinfo {author} {\bibfnamefont {P.~J.~D.}\ \bibnamefont
  {Crowley}}, \bibinfo {author} {\bibfnamefont {A.}~\bibnamefont {Datta}},
  \bibinfo {author} {\bibfnamefont {J.}~\bibnamefont {Nunn}}, \bibinfo {author}
  {\bibfnamefont {L.}~\bibnamefont {Zhang}}, \ and\ \bibinfo {author}
  {\bibfnamefont {I.}~\bibnamefont {Walmsley}},\ }\href {\doibase
  10.1103/PhysRevA.87.022313} {\bibfield  {journal} {\bibinfo  {journal} {Phys.
  Rev. A}\ }\textbf {\bibinfo {volume} {87}},\ \bibinfo {pages} {022313}
  (\bibinfo {year} {2013})}\BibitemShut {NoStop}%
\bibitem [{\citenamefont {Gioev}\ and\ \citenamefont
  {Klich}(2006)}]{Gioev2006Entanglement}%
  \BibitemOpen
  \bibfield  {author} {\bibinfo {author} {\bibfnamefont {D.}~\bibnamefont
  {Gioev}}\ and\ \bibinfo {author} {\bibfnamefont {I.}~\bibnamefont {Klich}},\
  }\href {\doibase 10.1103/PhysRevLett.96.100503} {\bibfield  {journal}
  {\bibinfo  {journal} {Phys. Rev. Lett.}\ }\textbf {\bibinfo {volume} {96}},\
  \bibinfo {pages} {100503} (\bibinfo {year} {2006})}\BibitemShut {NoStop}%
\bibitem [{\citenamefont {Shao}\ \textit {et~al.}(2014)\citenamefont {Shao},
  \citenamefont {Kim}, \citenamefont {Haldane},\ and\ \citenamefont
  {Rezayi}}]{Shao2014Entanglement}%
  \BibitemOpen
  \bibfield  {author} {\bibinfo {author} {\bibfnamefont {J.}~\bibnamefont
  {Shao}}, \bibinfo {author} {\bibfnamefont {E.-A.}\ \bibnamefont {Kim}},
  \bibinfo {author} {\bibfnamefont {F.~D.~M.}\ \bibnamefont {Haldane}}, \ and\
  \bibinfo {author} {\bibfnamefont {E.~H.}\ \bibnamefont {Rezayi}},\
  }\href@noop {} {\  (\bibinfo {year} {2014})},\ \Eprint
  {http://arxiv.org/abs/arXiv:1403.0577} {arXiv:1403.0577} \BibitemShut
  {NoStop}%
\bibitem [{\citenamefont {McHugh}\ \textit {et~al.}(2006)\citenamefont
  {McHugh}, \citenamefont {Ziman},\ and\ \citenamefont
  {Bu\ifmmode\check{z}\else\v{z}\fi{}ek}}]{McHugh}%
  \BibitemOpen
  \bibfield  {author} {\bibinfo {author} {\bibfnamefont {D.}~\bibnamefont
  {McHugh}}, \bibinfo {author} {\bibfnamefont {M.}~\bibnamefont {Ziman}}, \
  and\ \bibinfo {author} {\bibfnamefont {V.}~\bibnamefont
  {Bu\ifmmode\check{z}\else\v{z}\fi{}ek}},\ }\href {\doibase
  10.1103/PhysRevA.74.042303} {\bibfield  {journal} {\bibinfo  {journal} {Phys.
  Rev. A}\ }\textbf {\bibinfo {volume} {74}},\ \bibinfo {pages} {042303}
  (\bibinfo {year} {2006})}\BibitemShut {NoStop}%
\bibitem [{\citenamefont {Bronzan}(1988)}]{Bronzan1988ParamSU3}%
  \BibitemOpen
  \bibfield  {author} {\bibinfo {author} {\bibfnamefont {J.~B.}\ \bibnamefont
  {Bronzan}},\ }\href {\doibase 10.1103/PhysRevD.38.1994} {\bibfield  {journal}
  {\bibinfo  {journal} {Phys. Rev. D}\ }\textbf {\bibinfo {volume} {38}},\
  \bibinfo {pages} {1994} (\bibinfo {year} {1988})}\BibitemShut {NoStop}%
\bibitem [{\citenamefont {Bolukbasi}\ and\ \citenamefont
  {Dereli}(2006)}]{Bolukbasi2006qutrits}%
  \BibitemOpen
  \bibfield  {author} {\bibinfo {author} {\bibfnamefont {A.~T.}\ \bibnamefont
  {Bolukbasi}}\ and\ \bibinfo {author} {\bibfnamefont {T.}~\bibnamefont
  {Dereli}},\ }\href {http://stacks.iop.org/1742-6596/36/i=1/a=006} {\bibfield
  {journal} {\bibinfo  {journal} {J. Phys.: Conf. Series}\ }\textbf {\bibinfo
  {volume} {36}},\ \bibinfo {pages} {28} (\bibinfo {year} {2006})}\BibitemShut
  {NoStop}%
\bibitem [{\citenamefont {Anselmino}\ \textit {et~al.}(1993)\citenamefont
  {Anselmino}, \citenamefont {Predazzi}, \citenamefont {Ekelin}, \citenamefont
  {Fredriksson},\ and\ \citenamefont {Lichtenberg}}]{Anselmino_Diquarks}%
  \BibitemOpen
  \bibfield  {author} {\bibinfo {author} {\bibfnamefont {M.}~\bibnamefont
  {Anselmino}}, \bibinfo {author} {\bibfnamefont {E.}~\bibnamefont {Predazzi}},
  \bibinfo {author} {\bibfnamefont {S.}~\bibnamefont {Ekelin}}, \bibinfo
  {author} {\bibfnamefont {S.}~\bibnamefont {Fredriksson}}, \ and\ \bibinfo
  {author} {\bibfnamefont {D.~B.}\ \bibnamefont {Lichtenberg}},\ }\href
  {\doibase 10.1103/RevModPhys.65.1199} {\bibfield  {journal} {\bibinfo
  {journal} {Rev. Mod. Phys.}\ }\textbf {\bibinfo {volume} {65}},\ \bibinfo
  {pages} {1199} (\bibinfo {year} {1993})}\BibitemShut {NoStop}%
\end{thebibliography}%
\bibliographystyle{apsrev4-1}

\end{document}